\documentclass[%
 reprint,
 superscriptaddress,
 nofootinbib,
 nobibnotes,
 showpacs,
 amsmath,amssymb,
 aps,
 floatfix,
]{revtex4-2}

\ExplSyntaxOn
\msg_redirect_name:nnn { siunitx } { physics-pkg } { none }
\ExplSyntaxOff
\usepackage{silence}
\usepackage{graphicx}
\usepackage{dcolumn}
\usepackage{bm}
\usepackage[mathlines]{lineno}

\usepackage{xspace}
\usepackage{setspace}
\usepackage[version=4]{mhchem}
\usepackage{xfp}
\usepackage{circledsteps}
\usepackage{ulem}

\definecolor{revisiongreen}{RGB}{0,145,70}

\makeatletter
\AtEndDocument{\if@filesw\immediate\write\@auxout{\string\citation{REVTEX42Control}}\fi}
\makeatother

\renewcommand{\sout}{%
	\leavevmode
	\bgroup
	\markoverwith{\raise.55ex\hbox{\textcolor{red}{\rule{.2em}{1.0pt}}}}%
	\ULon
}

\pgfkeys{
  /csteps/inner xsep=2.5pt,
  /csteps/inner ysep=2.5pt,
  /csteps/outer color=black,
  /csteps/inner color=black,
}
\DeclareRobustCommand{\circled}[1]{{\CircledText{\scriptsize #1}}}

\usepackage{mathtools}

\usepackage[
  number-unit-product=~,
  per-mode=symbol,
  separate-uncertainty=true,
  range-units=single,
  range-phrase=\text{--},
  list-units=single,
  list-final-separator={\text{, and }}
]{siunitx}
\usepackage{physics}
\AtBeginDocument{\RenewCommandCopy\qty\SI} %

\usepackage[utf8]{inputenc}
\usepackage[T1]{fontenc}
\usepackage{times}
\usepackage{etoolbox}
\newcommand{\subm}[1]{_{\mathrm {#1}}}

\newcommand{\thetaK}{\theta\subm{K}}

\newcommand{\BKBO}{\ce{Ba_{1-x}K_xBiO3}\xspace}
\newcommand{\BKBOx}[1]{%
  \ce{Ba_{\fpeval{round(1-(#1), 2)}}K_{#1}BiO3}%
}
\newcommand{\Tc}{$T_{\mathrm{c}}$\xspace}

\usepackage[unicode,colorlinks=true,
  linkcolor=red, citecolor=red, urlcolor=black
]{hyperref}

\begin{document}


\title{Vortex pinning of \texorpdfstring{\BKBOx{0.38}}{Ba0.62K0.38BiO3} investigated by\texorpdfstring{\\}{ } magneto-optical Kerr-effect and magnetization measurements}

\author{Soichiro~Yamane}
\email{yamane.soichiro.a11@kyoto-u.jp}
\affiliation{Department of Electrical, Electronic, and Digital Science and Engineering, Kyoto University, Kyoto 615-8510, Japan}

\author{Sota~Nakamura}
\affiliation{Department of Electrical, Electronic, and Digital Science and Engineering, Kyoto University, Kyoto 615-8510, Japan}

\author{Atsutoshi~Ikeda}
\affiliation{Department of Electrical, Electronic, and Digital Science and Engineering, Kyoto University, Kyoto 615-8510, Japan}

\author{Dayu~Zhai}
\affiliation{School of Physics and Astronomy, University of Minnesota, Minneapolis, Minnesota 55455, USA}

\author{Siddarth~Gorregattu}
\affiliation{School of Physics and Astronomy, University of Minnesota, Minneapolis, Minnesota 55455, USA}

\author{Xing~He}
\affiliation{School of Physics and Astronomy, University of Minnesota, Minneapolis, Minnesota 55455, USA}

\author{Sudarshan~Sharma}
\affiliation{School of Physics and Astronomy, University of Minnesota, Minneapolis, Minnesota 55455, USA}

\author{Yipeng~Cai}
\affiliation{Department of Physics, Columbia University, New York, New York 10027, USA}

\author{Yasutomo~J.~Uemura}
\affiliation{Department of Physics, Columbia University, New York, New York 10027, USA}

\author{Martin~Greven}
\affiliation{School of Physics and Astronomy, University of Minnesota, Minneapolis, Minnesota 55455, USA}

\author{Shingo~Yonezawa}
\email{yonezawa.shingo.3m@kyoto-u.ac.jp}
\affiliation{Department of Electrical, Electronic, and Digital Science and Engineering, Kyoto University, Kyoto 615-8510, Japan}

\date{\today}

\begin{abstract}
	Vortex pinning plays a crucial role in determining properties of type-II superconductors.
	For example, it governs the irreversible magnetic response as well as dissipation caused by vortex motion.
	Here, we study vortex pinning in the three-dimensional oxide superconductor \BKBO using ultra-high-resolution magneto-optical Kerr effect (MOKE) and detailed magnetization measurements.
	We find that the zero-field MOKE signal in the superconducting state exhibits a pronounced magnetic-history dependence.
	This behavior closely resembles the remanent magnetization caused by trapped vortices.
	Furthermore, we demonstrate that the observed evolution of the MOKE signals is well described by Bean's critical-state model for trapped vortices.
	Our results establish MOKE as a viable optical and mesoscopic probe of vortex pinning in type-II superconductors, providing a new complementary approach to investigate mixed-state phenomena.
	We also find that the training-field dependence of the MOKE is linear near zero training field, without any anomalies indicative of spontaneous time-reversal-symmetry breaking in an unconventional superconducting state.
	Our study defines a clear protocol to distinguish vortex-induced MOKE responses from those associated with a time-reversal-symmetry broken superconducting order parameter.
\end{abstract}

\maketitle


\section{Introduction}\label{sec:intro}
	Vortex pinning is one of the most fundamental and practically important properties of type-II superconductors~\cite{Blatter1994-ve}.
	In the mixed state, magnetic flux penetrates the superconductor in the form of quantized vortices.
	Vortex pinning can lead to non-equilibrium phenomena such as hysteretic and remanent magnetization.
	From a fundamental science perspective, clarifying vortex-pinning behavior is important for understanding the superconducting mixed state, including the interplay between vortices and disorder~\cite{Bean1964-bt,Blatter1994-ve}.
	Furthermore, vortex motion causes energy dissipation, which is a fundamental issue in mixed-state physics as well as in superconducting applications~\cite{Eley2021-zm}.
	Controlling vortex pinning is therefore essential for reducing dissipation in superconducting devices and for realizing high-performance superconducting materials~\cite{Kwok2016-lc}.
	Information on vortex properties is also an important basis for the exploration of more exotic phenomena, such as chiral or nonunitary superconducting states with spontaneous time-reversal-symmetry breaking (TRSB)~\cite{Sigrist1991-nf,Kallin2016-wg,Ghosh2020-ky}.

	\BKBO (BKBO) is a perovskite-type oxide superconductor that has attracted considerable interest since its discovery in 1988~\cite{Cava1988-vn,Hinks1988-au,Sleight2015-wr,Griffitt2023-cp} [Fig.~\ref{fig:1}(a)].
	Superconductivity in BKBO emerges upon hole doping via K substitution.
	The compound exhibits a maximum superconducting critical temperature (\Tc) of about \SI{30}{\kelvin} near $x\sim0.4$~\cite{Cava1988-vn,Hinks1988-au,Griffitt2023-cp}.
	BKBO is known to be an archetypal type-II superconductor~\cite{Barilo1999-gl,Kumar1999-gz}.
	A roughly cubic Fermi surface centered at the $\Gamma$ point was first identified by positron angular-correlation measurements~\cite{Mosley1994-rj}.
	More recently, angle-resolved photoemission spectroscopy and electronic-structure calculations confirmed the simple three-dimensional low-energy electronic structure of BKBO~\cite{Wen2018-wi,Bhattacharyya2025-if} and revealed an isotropic superconducting gap~\cite{Wen2018-wi}.
	Thus, the low-energy electronic structure of BKBO is considerably simpler than those of many layered superconductors, such as cuprates~\cite{Damascelli2003-cw} and iron-based superconductors~\cite{Shibauchi2020-hf,Hirschfeld2011-iv}.
	Because of this electronic and structural simplicity as well as its relatively high \Tc, BKBO is a model system for the study of various superconducting phenomena, including vortex pinning.

	So far, vortex pinning in type-II superconductors has been studied mainly via magnetization and transport measurements~\cite{Bean1964-bt,Blatter1994-ve}.
	These techniques have been highly successful in revealing hysteresis, flux trapping, and critical-state behavior~\cite{Bean1964-bt,Blatter1994-ve}.
	However, they primarily probe bulk-averaged quantities, and do not directly access local magnetic responses produced by trapped vortices.
	On the other hand, real-space probes such as scanning tunneling microscopy, scanning
	superconducting quantum interference device (SQUID) microscopy, magnetic force microscopy, and magneto-optical imaging enable the visualization of individual vortices or local flux distributions~\cite{Hess1989-zw,Kirtley2010-ow,Jooss2002-oj,Goa2001-na}.
	A spatially selective mesoscopic probe is therefore desirable in order to bridge these two
	limits and obtain a complementary view of vortex pinning.

	The magneto-optical Kerr effect (MOKE) provides an alternative mesoscopic probe.
	In MOKE, the polarization plane of reflected light is rotated in the presence of magnetization~\cite{Kerr1877-tj}, and the Kerr angle $\thetaK$ is generally proportional to the local magnetization~\cite{Qiu2000-wi}.
	More broadly, MOKE has been used to investigate a variety of magnetic phenomena, including ferromagnetic and antiferromagnetic orders as well as nontrivial topological magnetic textures~\cite{Higo2018-yn,Kato2023-fp,Watanabe2026-sb,Yang2026-lc}.
	It has also been used as a probe of TRSB in unconventional superconductors~\cite{Kapitulnik2009-xh,Xia2006-rk,Schemm2014-cz,Wei2022-hn,Ajeesh2023-kz}.

	Because trapped vortices generate local magnetic fields, MOKE is naturally expected to be sensitive to vortex pinning.
	This technique probes magnetic properties within the optical spot, typically \SI{6}{\micro\meter} in diameter in our setup~\cite{Ikeda2026-bq,Yamane2026-jl}.
	Although this spot size does not allow individual vortices to be visualized, MOKE selectively probes the ensemble-averaged magnetic response from a restricted region of the sample.
	The measured Kerr signal is thus sensitive to the local remanent field generated by trapped vortices.
	This mesoscopic spatial selectivity distinguishes MOKE from bulk magnetization measurements and allows the remanent vortex response in a chosen region of the sample to be directly compared with mesoscopic pinning models.
	In addition, due to the contactless nature of the measurement, it can probe magnetic responses without complications associated with transport measurements, such as attaching contact leads.
	Despite this potential, direct experimental demonstrations of vortex pinning detected by MOKE remain limited~\cite{Wei2022-hn,Ajeesh2023-kz}.
	The relation between MOKE signals and vortex pinning therefore remains insufficiently established, particularly in comparison with bulk magnetization measurements.

	Here we report an ultra-high-resolution MOKE study of vortex pinning in BKBO, combined with precise bulk magnetization measurements.
	In the MOKE measurements, using a newly developed movable magnetic shield to suppress remnant fields, we realized a zero-field environment for MOKE measurements while retaining the ability to apply magnetic fields up to several teslas before the measurements
	We show that the MOKE signal exhibits clear magnetic-history dependence.
	In particular, MOKE signals measured during zero-field warming (ZFW) after field cooling (FC) reveal a remanent response that depends on the training field applied during cooling.
	This MOKE behavior can be ascribed to small magnetic fields caused by trapped vortices, as confirmed by very similar behavior observed in magnetization measurements.
	The remanent response is explained by Bean's critical-state model~\cite{Bean1964-bt}, which describes the coarse distribution of magnetic flux in a superconductor with strong pinning.
	We also found that the training-field dependence of the MOKE signal is linear near zero training field, with no additional low-field anomalies attributable to TRSB in the superconducting order parameter.
	These results establish MOKE as a viable optical probe of vortex pinning and demonstrate that MOKE can directly access vortex-related phenomena.
	Our study also provides a basis for reliable protocols for detecting TRSB in unconventional superconductors.

\section{Methods}\label{sec:methods}

	\begin{figure}[t]
		\centering
		\includegraphics[width=246pt, bb=0.000000 0.000000 246.00000 187.199993]{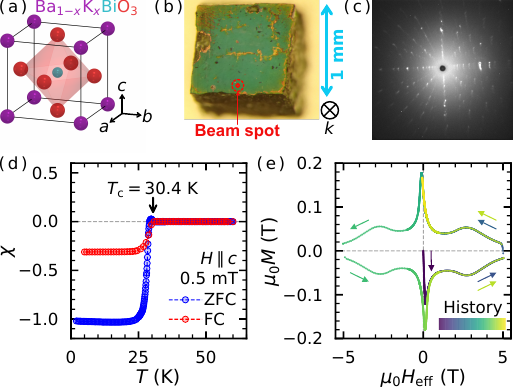}
		\caption{
			(a) Pseudocubic subcell of \BKBO derived from the tetragonal $I4/mcm$ structure.
			The \ce{BiO6} octahedron is indicated by the pink faces.
			The crystal structure was rendered using pyvista and pymatgen packages~\cite{Sullivan2019-up,Ong2013-hn}.
			(b) Photograph of the \BKBOx{0.38} single-crystal sample used in this study.
			The wave vector $\bm{k}$ of the incident light in the MOKE measurements is perpendicular to this surface, as indicated in the figure.
			The approximate position of the optical spot in our MOKE study is depicted with red circles.
			(c) Laue diffraction pattern of the sample, confirming the high quality of the single crystal.
			The studied surface is confirmed to be the $(001)$ plane in the pseudocubic notation.
			(d) Temperature dependence of the volumetric magnetic susceptibility $\chi$ (dimensionless in the SI units) of the single-crystal sample measured under zero-field-cooling (ZFC) and field-cooling (FC) conditions under an applied magnetic field corresponding to $\mu_0H=\SI{0.5}{\milli\tesla}$ applied perpendicularly to the $(001)$ surface in (b).
			The demagnetization correction for $\chi$ is taken into account.
			The onset critical temperature \Tc of \SI{30.4}{\kelvin} is indicated by the arrow.
			(e) Magnetization $M$ as a function of the effective applied magnetic field $H_{\rm eff}$ at \SI{5}{\kelvin} after ZFC.
			The hysteresis loop is caused by vortex pinning in the superconducting state.
			The sweeping directions of the magnetic field are indicated by arrows and highlighted by the color of the data points.
			$H_{\rm eff}$ is the magnetic field corrected using the demagnetization factor of the sample.
		}\label{fig:1}
	\end{figure}

	\begin{figure}[t]
		\centering
		\includegraphics[width=246pt, bb=0.000000 0.000000 246.000001 340.080001]{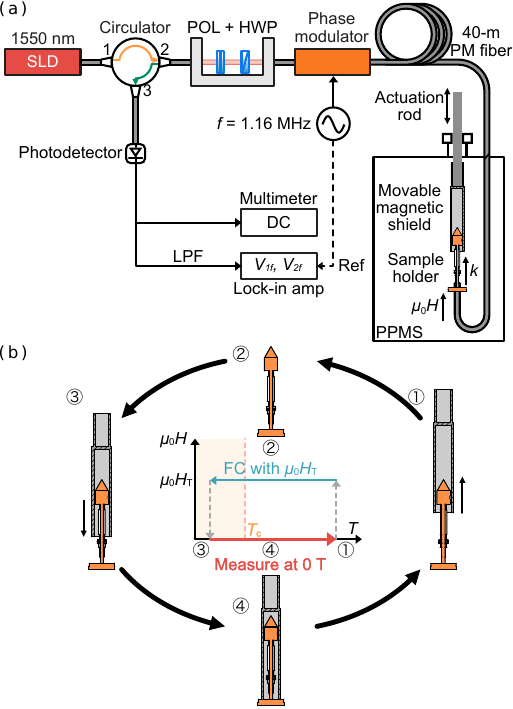}
		\caption{
			(a) Schematic of the polar MOKE measurement setup, where the incident and reflected beams are collinear with the applied magnetic field.
			The sample is placed inside a commercial $^4$He cryostat (PPMS).
			The movable magnetic shield (permalloy) is placed around the sample to suppress remnant fields during the zero-field MOKE measurements.
			The detailed design of the sample fixture and optical path is described in Ref.~\cite{Ikeda2026-bq}.
			(b) Field-training protocol used in this MOKE study.
			To investigate the vortex pinning,
			we employed the following field-training protocol:
			\circled{1}
			The shield was moved away from the magnetic-field region
			to avoid magnetizing the permalloy shield,
			and the magnetic field was set to a target training field
			$H_{\rm T}$ in the normal state above \Tc.
			\circled{2}
			The sample was cooled to a target temperature $T$
			(typically \SI{5}{\kelvin})
			under the training field $H_{\rm T}$.
			\circled{3}
			The applied field was then set to zero, after which the shield was placed back to around the sample space.
			\circled{4}
			Zero-field MOKE measurements are performed during warming above \Tc.
		}\label{fig:2}
	\end{figure}

	The BKBO single crystal used in this study was grown by an electrochemical method proposed in Ref.~\citenum{Griffitt2023-cp}; see also Refs.~\citenum{Norton1989-ye,Barilo1999-gl,Nishio2001-xg}.
	In order to obtain a fresh and clean surface for the optical measurement, we polished the sample surface $(001)$ with diamond slurry with particle sizes of \SI{0.1}{\micro\meter} and polishing films with grit sizes down to \SI{0.02}{\micro\meter}.
	The sample dimensions are approximately \SI{0.95}{\milli\meter}~$\times$~\SI{0.94}{\milli\meter}~$\times$~\SI{0.74}{\milli\meter} as shown in Fig.~\ref{fig:1}(b).
	The mass of the sample is \SI{6.090}{\milli\gram}.
	The crystal quality was confirmed by Laue diffraction photographs taken with a Laue camera (Rigaku, RASCO-BL2).
	The sharp and well-defined spots in the Laue pattern [Fig.~\ref{fig:1}(c)] confirm the high quality of the single crystal.
	Magnetization and MOKE measurements were carried out using the same sample.

	Magnetization measurements were performed using a commercial DC SQUID magnetometer (Quantum Design, MPMS-XL).
	The superconductivity of the sample was characterized by measuring the temperature dependence of the volumetric magnetic susceptibility $\chi$.
	Here, $\chi$ is calculated as $\chi = M / H_{\rm eff}$, where $M$ is the volumetric magnetization and $H_{\rm eff} = H_{\rm ext} - NM$ is the effective magnetic field corrected for demagnetization effects.
	In this expression, $H_{\rm ext}$ is the externally applied magnetic field, and $N$ is the demagnetization factor of the sample.
	For fields perpendicular to the $(001)$ surface, the demagnetization factor is evaluated to be approximately $N = 0.39$ using the method described in Ref.~\citenum{Aharoni1998-sx}.
	\Tc is determined to be \SI{30.4}{\kelvin} from the onset of the diamagnetic response in the zero-field cooling (ZFC) data, as indicated by the arrow in Fig.~\ref{fig:1}(d).
	The doping level of the sample was estimated to be $x = 0.38$ from the measured \Tc.
	This estimate is based on both our empirical \Tc-doping relation obtained from electron probe microanalysis and inductively coupled plasma measurements on BKBO samples grown using the same method, and is consistent with the previously reported \Tc-doping relation in Ref.~\citenum{Pei1990-xo}.
	Figure~\ref{fig:1}(e) displays the clear hysteresis loop, which indicates strong vortex pinning in the sample, consistent with previous studies of BKBO~\cite{Kim2000-sy,Tao2015-sm}.

	MOKE measurements were performed using a zero-area-loop (loop-less) fiber-optic Sagnac interferometer working at a wavelength of \SI{1550}{\nano\meter}~\cite{Xia2006-sg,Ikeda2026-bq,Xia2006-rk}, as described in Fig.~\ref{fig:2}(a).
	The interferometer is constructed based on the design reported in Ref.~\cite{Xia2006-sg}, but in our implementation all free-space optical components were replaced with fiber-optic components to improve usability and stability for MOKE measurements.
	Incoherent light from a superluminescent diode (Thorlabs, S5FC1550S-A2) with a center wavelength of \SI{1550}{\nano\meter} (\SI{0.8}{\eV}) and a bandwidth of \SI{90}{\nano\meter} is used.
	The power of the incident light to the sample is typically set to be around \SI{550}{\micro\watt}, corresponding to \SI{400}{\milli\ampere} to the diode in the light source.
	The light is delivered to the Sagnac interferometer through a polarization-maintaining (PM) optical fiber.
	Inside the interferometer, the beam is split into the fast and slow axes of the PM fiber by a half-wave plate embedded in a fiber bench (Optoquest corporation, PCUA-15-P/FA).
	The counter-propagating beams pass through a phase modulator driven at $f=\SI{1.1589}{\mega\hertz}$.
	This frequency was determined by fitting the modulation-frequency dependence of the second harmonic and DC signals.
	After passing through the phase modulator and \SI{40}{\meter} of PM fiber, the beams are focused to the sample surface by a gradient-index lens (Photonic Science Technology)~\cite{Ikeda2026-bq}.
	A polyimide-based quarter-wave plate (NTT Advanced Technology Corporation, AT-QWP-4A) was placed on the surface of the lens to convert linearly polarized lights to circularly polarized lights and vice versa.
	Circularly polarized lights are focused onto the sample surface, forming an optical spot of approximately \SI{6}{\micro\meter}.
	In the present measurements, the optical spot was positioned approximately \SI{0.1}{\milli\meter} from the nearest sample edge as shown in Fig.~\ref{fig:1}(b).
	The alignment of the incident and reflected beams is facilitated by our sample fixture, which enables compact and reproducible optical-path adjustment in the restricted sample space~\cite{Ikeda2026-bq}.
	Non-reciprocal optical phase shift upon reflection from the sample surface is induced by the Kerr effect, and the two circularly polarized lights return to the interferometer through the same optical path.
	The interference signal is detected by a photodetector.
	Voltage signals corresponding to the first and second harmonics of the modulation frequency, $V_{1f}$ and $V_{2f}$, are extracted using a dual-frequency lock-in amplifier (NF corporation, LI5660).
	The Kerr angle $\thetaK$ is calculated from these signals using the relation $\thetaK = (1/2) \tan^{-1} \left[ J_2(2\phi_m) V_{1f} / J_1(2\phi_m) V_{2f} \right]$~\cite{Xia2006-sg}.
	Here, $J_1$ and $J_2$ are the first- and second-order Bessel functions of the first kind, and $\phi_m$ is the modulation amplitude of the phase modulator.
	Here, $\phi_m$ is set to be approximately \SI{0.92}{\radian} to maximize the sensitivity of the first-harmonic detection.
	A low-pass filter (LPF, Thorlabs, EF508, \SI{-3}{\decibel} rejection at \SI{1.35}{\mega\hertz}) was placed before the lock-in amplifier to reduce the large $V_{2f}$ signal, resulting in a resolution of the order of nanoradian.

	Cryogenic MOKE measurements were conducted using a commercial $^4$He cryostat (Quantum Design, PPMS).
	The cryostat is equipped with a superconducting magnet capable of generating magnetic fields up to \SI{9}{\tesla}.
	MOKE was measured using the polar MOKE setup, in which incident and reflected beams are collinear with the applied magnetic field.
	This configuration detects magnetization perpendicular to the sample surface.

	With a superconducting magnet, remnant magnetic fields often remain even after setting the nominal field to zero~\cite{Buchner2018-xm}.
	To eliminate the influence of remnant fields, we installed a movable magnetic shield made of permalloy around the sample space, which effectively attenuates the residual field.
	When an external magnetic field is applied, the shield is placed away from the sample space.
	This avoids magnetizing permalloy and distorting the applied field, while the sample remains inside the bore of the superconducting magnet [Fig.~\ref{fig:2}].
	Details of this measurement system with the movable shield will be described elsewhere.

	We comment on the sign of the Kerr angle.
	In the present design of our MOKE sample holder, the magnetic field is parallel to the wavevector of the incident light [Fig.~\ref{fig:2}(a)].
	This geometry is opposite to the ordinary configuration, where the magnetic field and the wavevector are antiparallel.
	To have the sign of $\theta_{\rm K}$ that is consistent with the one obtained in the conventional configuration, we flip the sign of the as-measured Kerr angle.
	This convention should be valid because BKBO has a globally centrosymmetric crystal structure~\cite{Pei1990-xo,Sleight2015-wr,Griffitt2023-cp}, although inversion symmetry is broken locally~\cite{Griffitt2023-cp}.
	We have also checked the sign and magnitude of $\thetaK$ in our convention using magnetite (\ce{Fe3O4}),
	which exhibits a known saturated Kerr angle of approximately \SI{-0.54}{\milli\radian} at
	\SI{1550}{\nano\meter} under a positive magnetic field~\cite{Ikeda2026-bq,Yamane2026-jl,Simsa1980-us,Fontijn1997-hr}.

\section{Results and Discussion}\label{sec:results_and_discussion}

	\begin{figure*}[ht]
		\centering
		\includegraphics[width=510pt, bb=0.000000 0.000000 510.000006 313.000002]{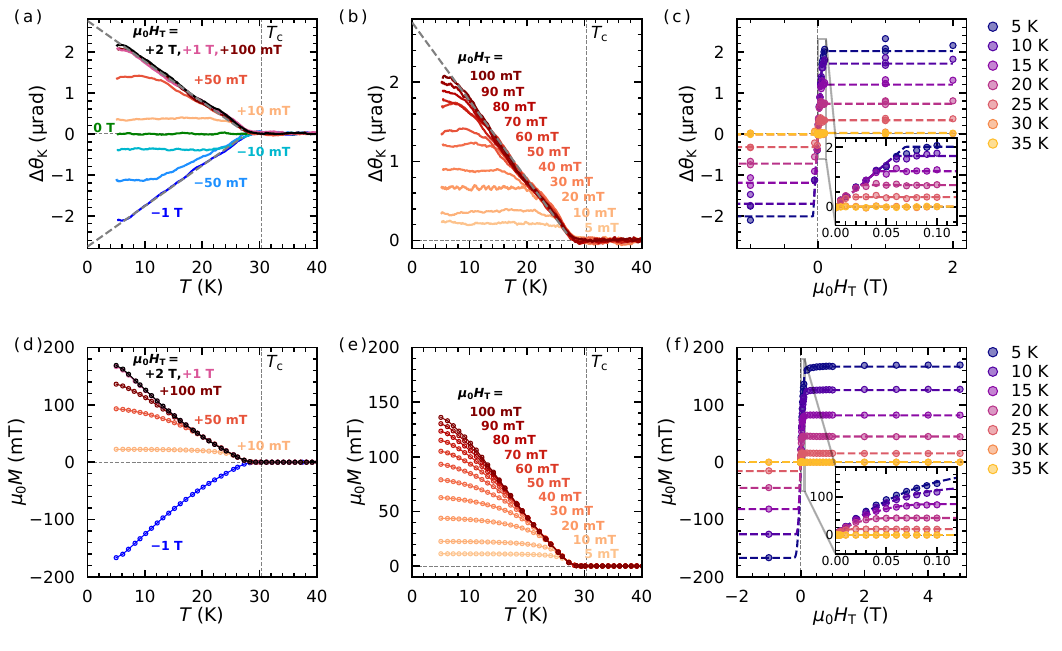}
		\caption{
			(a) Temperature dependence of the Kerr angle change $\Delta\theta_{\rm K}$ during ZFW after FC under various training fields $H_{\rm T}$.
			Here, $\Delta\theta_{\rm K}(T)$ is defined as $\theta_{\rm K}(T)-\bar{\theta}_{\rm K}(T>35\,{\rm K})$.
			The second term is the averaged Kerr angle in the normal state and is subtracted as a background offset.
			The broken lines in panels (a) and (b) are guides to the eye that illustrate the saturation behavior, obtained from linear fits to the high-$H_{\rm T}$ data.
			(b) Detailed training-field dependence of the remanent Kerr angle $\Delta\theta_{\rm K}$ measured with finer steps of $H_{\rm T}$.
			(c) Training-field dependence of the Kerr angle change $\Delta\theta_{\rm K}$ at various temperatures.
			The dashed lines show fits using the local clipped-linear saturation function described by Eq.~\eqref{eq:clipped_linear}.
			The inset expands the low-field region.
			(d) Temperature dependence of the magnetization $M$ measured during ZFW after FC under various training fields $H_{\rm T}$.
			The data show magnetic-history dependence consistent with the MOKE results.
			(e) Detailed training-field dependence of the remanent magnetization $M$ with finer steps of $H_{\rm T}$.
			(f) Training-field dependence of the magnetization $M$ at various temperatures.
			The dashed lines show fits using the bulk-averaged Bean model described by Eq.~\eqref{eq:bean_gradient}.
		}\label{fig:3}
	\end{figure*}

	\subsection{Results of experiments}

		In order to investigate vortex pinning in BKBO, we performed MOKE measurements under various field-training conditions.
		The field-training protocol is summarized in Fig.~\ref{fig:2}(b).
		First, the sample was cooled through \Tc to a target temperature $T$ (typically \SI{5}{\kelvin}) under a training field $H_{\rm T}$.
		Then, the applied field was set to zero, and the Kerr angle was measured during ZFW above \Tc.
		During the ZFW process, the movable magnetic shield was placed around the sample to suppress artifacts caused by remnant fields.

		Figure~\ref{fig:3}(a) shows the temperature dependence of the Kerr angle change $\Delta\theta_{\rm K}$ measured during ZFW after FC under various training fields.
		Here, the Kerr angle change $\Delta\theta_{\rm K}$ is defined as
		\begin{equation}
			\Delta\theta_{\rm K}(T)
			=
			\theta_{\rm K}(T)
			-
			\bar{\theta}_{\rm K}(T>35\,{\rm K}),
		\end{equation}
		where $\bar{\theta}_{\rm K}(T>35\,{\rm K})$ is the averaged Kerr angle above \SI{35}{\kelvin}.
		This subtraction removes a constant offset determined in the normal state.
		A clear remanent Kerr response is observed only below \Tc, as shown in Fig.~\ref{fig:3}(a).
		The sign of $\Delta\theta_{\rm K}$ depends on the polarity of the training field $H_{\rm T}$.
		These observations show that the remanent Kerr signal is a superconductivity-related, field-history-dependent magnetic response.
		The magnitude of the low-temperature remanent response increases with increasing $|H_{\rm T}|$ and approaches saturation for $|\mu_0H_{\rm T}| \gtrsim \SI{0.1}{\tesla}$.
		To show this saturation behavior more clearly, we plot results of linear fits to the high-$H_{\rm T}$ data with the broken curves in Fig.~\ref{fig:3}(a) and Fig.~\ref{fig:3}(b).
		For $|\mu_0 H_{\rm T}| \lesssim \SI{0.1}{\tesla}$, $\Delta\theta_{\rm K}$ exhibits a pronounced plateau-like behavior up to a characteristic temperature, above which $\Delta\theta_{\rm K}$ merges with the curves obtained at higher training fields.
		The detailed training-field dependence measured with finer field steps is shown in Fig.~\ref{fig:3}(b), where the monotonic evolution of the remanent Kerr response is more clearly resolved.
		Since $\Delta\theta_{\rm K}$ measures the local magnetic response in general, this monotonic increase in $\Delta\theta_{\rm K}$ with increasing $H_{\rm T}$ indicates an increase in the local flux retained after the training process.

		Figure~\ref{fig:3}(c) summarizes the training-field dependence of $\Delta\theta_{\rm K}$ at selected temperatures.
		Below \Tc, $\Delta\theta_{\rm K}$ changes approximately linearly near zero training field and approaches a saturated value at larger $|H_{\rm T}|$.
		Above \Tc, $\Delta\theta_{\rm K}$ remains close to zero and shows no clear dependence on $H_{\rm T}$.

		To compare the MOKE response with the bulk magnetic response,
		we performed magnetization measurements on the same sample using the same field-training protocol.
		Figures~\ref{fig:3}(d) and~\ref{fig:3}(e) show the temperature dependence of $M$ measured during ZFW after FC under various $H_{\rm T}$.
		A remanent magnetization is observed below \Tc, and its sign is determined by the polarity of $H_{\rm T}$.
		The remanent magnetization decreases with increasing temperature and vanishes above \Tc.
		The MOKE and magnetization data resemble each other, although the plateau-like behavior is less pronounced in $M(T)$.
		It has been known that the remanent magnetization after field cooling is a conventional bulk signature of vortex pinning in type-II superconductors~\cite{Bean1964-bt,Blatter1994-ve,Kim2000-sy,Tao2015-sm}.
		The qualitative similarity between the temperature dependences of $M$ and $\Delta\theta_{\rm K}$ indicates that the MOKE response also originates from trapped vortices.
		Figure~\ref{fig:3}(f) shows the training-field dependence of $M$ at selected temperatures.
		The shapes of the $M(H_{\rm T})$ curves are again very similar to those of $\Delta\theta_{\rm K}(H_{\rm T})$: both responses change sign depending on the polarity of $H_{\rm T}$, increase nearly linearly at low fields, and approach saturation at higher fields.
		This close correspondence between the established bulk pinning response and the local Kerr response provides strong evidence that the remanent Kerr signal is governed by vortex pinning.

	\subsection{Analysis based on the critical-state model}

		The critical-state model provides a simple phenomenological description of irreversible magnetization in type-II superconductors with vortex pinning~\cite{Bean1964-bt,Kim1963-xe,Blatter1994-ve}.
		The model treats the vortex distribution in a coarse-grained manner, replacing discrete vortices by a continuous magnetic-flux-density profile.
		In this model, vortices that penetrate into the sample are pinned by defects.
		When the external magnetic field is varied, vortices move at a rate limited by the critical current density $j_{\rm c}$.
		In combination with Ampere's law, the magnetic-flux-density distribution $\bm{B}(\bm{r})$ in the critical state is described by
		\begin{equation}
			\nabla_{\bm{r}} \times \bm{B}(\bm{r})
			=
			\mu_0\bm{j}(\bm{r}),
			\qquad
			|\bm{j}(\bm{r})|
			=
			j_{\rm c}.
		\end{equation}

		Bean further assumed that $j_{\rm c}$ is independent of the magnetic field~\cite{Bean1964-bt}.
		This results in linear dependence of $\bm{B}(\bm{r})$ on the spatial coordinate $\bm{r}$, with a slope determined by $j_{\rm c}$.
		When the applied field is removed after field cooling, this critical-state profile can remain in the sample, producing remanent magnetization and hysteresis.
		We define the full critical-state penetration field $H_{\rm p}$ as follows:
		\begin{equation}
			\mu_0H_{\rm p}
			\sim
			\mu_0j_{\rm c}L.
		\end{equation}
		Here, $H_{\rm p}$ is the field scale at which the Bean-like linear flux distribution reaches the center of the sample with the radius $L$.
		This model provides a simple framework for comparing the bulk magnetization response with the local remanent MOKE response.

		\begin{figure*}[ht]
			\centering
			\includegraphics[width=510.0pt, bb=0.000000 0.000000 510.000006 305.039999]{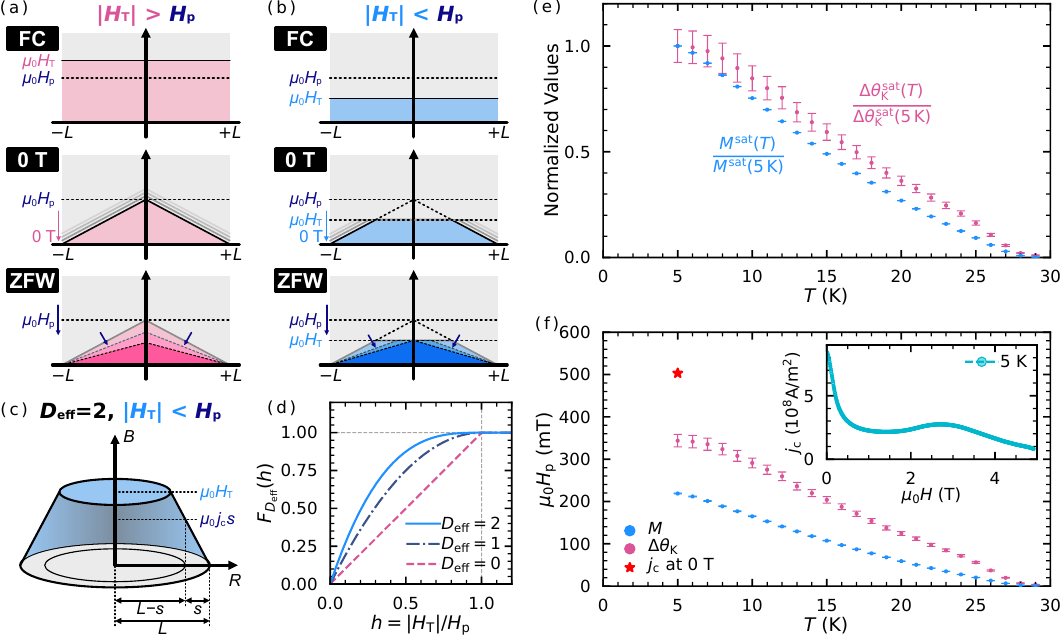}
			\caption{
			(a)(b) Magnetic-flux distribution $\bm{B}(\bm{r})$ arising from vortex pinning, as described by Bean's critical-state model~\cite{Bean1964-bt}, for the field-training protocol used in this study.
			Here, $H_{\rm p}$ denotes the full flux penetration field in the phenomenological Bean-model picture.
			(a) When the training field satisfies $|H_{\rm T}|\gtrsim H_{\rm p}$, the final flux distribution $\bm{B}(\bm{r})$ at zero field is expected to be independent of $H_{\rm T}$.
			Upon warming, the reduction of $j_{\rm c}$ leads to a uniform decrease of the slope of the linear $B(\bm{r})$ profile.
			This leads to $H_{\rm T}$-independent remanent response as a function of temperature.
			(b) When $|H_{\rm T}|<H_{\rm p}$, a region with constant $B(\bm{r})$ remains around the center of the sample.
			As the temperature increases, the decrease in $j_{\rm c}$ modifies the flux distribution primarily near the sample edge, while leaving the central region nearly unchanged.
			This leads to the plateau-like response with a smaller training field scale at low temperatures, as observed in our measurements.
			(c) Schematic of the remnant flux distribution with a low training field in a cylindrical sample.
			The linear $B(\bm{r})$ profile extends only over the relevant length scale with a gradient set by the critical current density $j_{\rm c}$.
			(d) Saturation functions $F_{D_{\rm eff}}(h)$ for various effective dimensions $D_{\rm eff}$.
			The function $F_{D_{\rm eff}}(h)$ describes the training-field dependence of the remanent response in the Bean-model picture.
			(e) Temperature dependence of the saturation parameters $R^{\rm sat}$ obtained from the magnetization and MOKE measurements.
			The vertical axis displays the normalized values of $M^{\rm sat}(T)$ and $\Delta\theta_{\rm K}^{\rm sat}(T)$, where each parameter is normalized by its value at \SI{5}{\kelvin}.
			Both parameters show a similar temperature dependence.
			(f) Full flux penetration field $H_{\rm p}$ obtained from different probes plotted as a function of temperature.
			The inset shows the magnetic-field dependence of the effective critical current density $j_{\rm c}(H)$ at \SI{5}{\kelvin}, estimated from the magnetization hysteresis loop using critical-state analysis~\cite{Kim1963-xe,Bean1964-bt,Chen1990-yo,Kim2000-sy,Tao2015-sm}.
			The comparison indicates that these field scales are reasonably consistent, given the simplified nature of the critical-state analysis.
			}\label{fig:4}
		\end{figure*}

		Figure~\ref{fig:4} summarizes the Bean-model interpretation and the fitting procedures used to analyze the remanent response.
		Figures~\ref{fig:4}(a) and~\ref{fig:4}(b) illustrate the basic critical-state picture for the field-training protocol.
		When the training field satisfies $|H_{\rm T}|\gtrsim H_{\rm p}$, the Bean-like profile after zeroing the external field reaches the center of the sample and the remanent vortex distribution is expected to be independent of $H_{\rm T}$ [Fig.~\ref{fig:4}(a)].
		Upon warming, the reduction of $j_{\rm c}$ uniformly decreases the slope of this linear $B(\bm{r})$ profile.
		This leads to the overlap of the remanent response as a function of temperature for different higher training fields.
		As a result, the temperature-dependent remanent responses for different high training fields collapse onto a common curve.
		In contrast, for $|H_{\rm T}|<H_{\rm p}$, the linear $\bm{B}(r)$ distribution is only partially developed [Fig.~\ref{fig:4}(b)].
		Around the center of the sample, there remains a region with $B(\bm{r}) = \mu_0 H_{\rm T}$.
		In this case, the remanent response depends strongly on $H_{\rm T}$.
		At low temperatures, the decrease in $j_{\rm c}$ modifies the flux distribution mainly near the sample edge, while the central region maintains the constant $B(\bm{r})$ profile.
		This produces a plateau-like temperature dependence of the remanent response for low training fields.
		With further warming, $j_{\rm c}(T)$ is reduced and the linear $B(\bm{r})$ profile eventually reaches the center of the sample.
		As a result, the remanent response merges with the high-$H_{\rm T}$ curves at certain temperatures and vanishes near \Tc.
		This simple scenario qualitatively explains the observed behavior of $\Delta\theta_{\rm K}$ and $M$.

		We next formulate this picture as equations for further quantitative analysis.
		Figure~\ref{fig:4}(c) schematically shows the two-dimensional unsaturated remanent profile $\bm{B}(\bm{r})$ in a cylindrical sample that is used as the starting point for the model.
		The difference between MOKE and magnetization measurements can be viewed as a consequence of different spatial averages of this Bean-like remanent-flux profile.
		More generally, the remanent response measured by a probe $i$ can be expressed as a spatial integration of the remanent magnetic-flux-density profile,
		\begin{equation}
			R_i(H_{\rm T},T)
			=
			A_i
			\int
			W_i(\bm{r})
			B_{\rm rem}(\bm{r};H_{\rm T},T)
			d^3r,
			\label{eq:response_functional}
		\end{equation}
		where $R_i$ represents either $\Delta\theta_{\rm K}$ or $M$, $A_i$ is a probe-dependent conversion factor, $W_i(\bm{r})$ is a probe-dependent spatial weighting function, and $B_{\rm rem}$ is the remanent magnetic flux density profile after removing a training field $H_{\rm T}$.
		To derive the probe-dependent training-field dependences, we consider a simplified Bean profile expressed below.
		In a field-cooled remanent state, the local remanent flux density is approximated as
		\begin{equation}
			B_{\rm rem}(s;H_{\rm T},T)
			=
			\operatorname{sgn}(H_{\rm T})
			\min\left[
			\mu_0|H_{\rm T}|,
			\mu_0j_{\rm c}(T)s
			\right], \label{eq:bean_profile}
		\end{equation}
		where $s$ is the distance from the nearest sample edge.
		This form reflects the Bean critical-state gradient $dB/ds\simeq\mu_0j_{\rm c}$.

		For a spatially averaged response such as bulk magnetization, we approximate the weighting function as uniform, $W_{\rm bulk}(\bm{r})\simeq 1/V$, where $V$ is the sample volume.
		Assuming homogeneity along the field direction, the three-dimensional volume element can be reduced to an effective one-dimensional form along the edge-distance coordinate,
		\begin{equation}
			\int d^3r
			\rightarrow
			z_0\int ds\,G_{D_{\rm eff}}(s),
			\qquad
			G_{D_{\rm eff}}(s)
			\propto
			(L-s)^{D_{\rm eff}-1}.
		\end{equation}
		Here, $G_{D_{\rm eff}}(s)$ is a simplified geometric weight obtained by collecting the contributions from regions with the same $s$, and $z_0$ is the sample dimension along the field direction.
		With this approximation, the spatial average for $D_{\rm eff}\geq1$ can be written as
		\begin{align}
			R_{D_{\rm eff}}(H_{\rm T},T)
			 & \propto
			\operatorname{sgn}(H_{\rm T})
			\int_0^L
			G_{D_{\rm eff}}(s) \notag \\
			 & \quad \times
			\min\left[
			\mu_0|H_{\rm T}|,
			\mu_0j_{\rm c}(T)s
			\right]ds.
		\end{align}
		For example, $D_{\rm eff}=2$ yields $G_{D_{\rm eff}}(s)\propto L-s$, corresponding to a model for a cylindrical superconductor.
		Using scaled distance $t=s/L$ and bulk-normalized field $h_b=|H_{\rm T}|/H_{\rm p}(T)$, the field-dependent part of the bulk-averaged integral for $h_{\rm b}\leq1$ is proportional to
		\begin{align}
			I_{D_{\rm eff}}(h_{\rm b})
			 & =
			\int_0^{h_{\rm b}} t(1-t)^{D_{\rm eff}-1} dt
			+
			\int_{h_{\rm b}}^1 h_{\rm b}(1-t)^{D_{\rm eff}-1} dt \\
			 & =
			\frac{1-(1-h_{\rm b})^{D_{\rm eff}+1}}{D_{\rm eff}(D_{\rm eff}+1)},
		\end{align}
		while the saturated value is $I_{D_{\rm eff}}(1)=1/[D_{\rm eff}(D_{\rm eff}+1)]$.
		This result motivates us to define a normalized saturation function with a generic argument $h$ as $F_{D_{\rm eff}}(h)=I_{D_{\rm eff}}(h)/I_{D_{\rm eff}}(1)$.

		The local limit $D_{\rm eff}=0$, corresponding to MOKE, is obtained using $W_{\rm local}(\bm{r}) \simeq \delta(\bm{r}-\bm{r}_{\rm spot})$, where $\bm{r}_{\rm spot}$ is the probed position.
		For a spot located at an effective distance $s_{\rm spot}$ from the nearest sample edge, the local response is governed by the saturation field defined as $H_{\rm spot}(T)\sim j_{\rm c}(T)s_{\rm spot}$.
		Using the corresponding normalized field $h_{\rm spot}=|H_{\rm T}|/H_{\rm spot}(T)$,
		we obtain $F_0(h_{\rm spot})=\min(h_{\rm spot},1)$, which is identical to $1-(1-h)^{D_{\rm eff}+1}$ with $D_{\rm eff}=0$ for $0\leq h\leq1$.

		Overall, the same normalized functional form can be used for both local and spatially averaged responses, provided that its argument is normalized by the appropriate field scale:
		\begin{equation}
			F_{D_{\rm eff}}(h)
			=
			\begin{cases}
				1-(1-h)^{D_{\rm eff}+1}, & 0\leq h\leq1, \\
				1,                       & h\geq1.
			\end{cases}
		\end{equation}
		The shapes of this function are shown in Fig.~\ref{fig:4}(d).
		The case $D_{\rm eff}=0$ corresponds to a local sampling of the Bean profile, whereas $D_{\rm eff}=2$ corresponds to averaging over a two-dimensional cross section.
		With increasing $D_{\rm eff}$, the saturation curve becomes more strongly convex and approaches saturation more rapidly at low $h$.

		For a bulk-averaged response, the remanent response can then be written as
		\begin{equation}
			R_{D_{\rm eff}}(H_{\rm T},T)
			=
			R_{D_{\rm eff}}^{\rm sat}(T)
			\operatorname{sgn}(H_{\rm T})
			F_{D_{\rm eff}}\left(\frac{|H_{\rm T}|}{H_{\rm p}(T)}\right),
			\label{eq:bean_saturation_response}
		\end{equation}
		where $R_{D_{\rm eff}}^{\rm sat}(T)$ is the saturated remanent response at temperature $T$.
		For magnetization, the total magnetization is proportional to the integration of the nonuniform Bean profile over the cross section perpendicular to the applied field.
		Although the actual sample shape is nearly cubic [Fig.~\ref{fig:1}(b)],
		we adopt the cylindrical model ($D_{\rm eff} = 2$) as a simple approximation for the bulk-averaged response.
		Therefore, we have
		\begin{equation}
			M(H_{\rm T},T)
			=
			M^{\rm sat}(T)
			\operatorname{sgn}(H_{\rm T})
			F_2\left(\frac{|H_{\rm T}|}{H_{\rm p}(T)}\right).\label{eq:bean_gradient}
		\end{equation}
		For MOKE, we set $D_{\rm eff}=0$ and use $h_{\rm spot}$ to obtain
		\begin{equation}
			\Delta\theta_{\rm K}(H_{\rm T},T)
			=
			\Delta\theta_{\rm K}^{\rm sat}(T)
			\operatorname{sgn}(H_{\rm T})
			F_0\left(\frac{|H_{\rm T}|}{H_{\rm spot}(T)}\right).\label{eq:clipped_linear}
		\end{equation}

		The dashed curves in Figs.~\ref{fig:3}(c) and~\ref{fig:3}(f) show fits using these saturation functions.
		The successful fits using these simple model functions show that the local form with $D_{\rm eff}=0$ reproduces the MOKE data, while the bulk-averaged form with $D_{\rm eff}=2$ captures the magnetization data.
		The extracted fitting parameters are summarized in Figs.~\ref{fig:4}(e) and~\ref{fig:4}(f).
		Figure~\ref{fig:4}(e) shows the normalized saturation values, $\Delta\theta_{\rm K}^{\rm sat}(T)/\Delta\theta_{\rm K}^{\rm sat}(5\,{\rm K})$ and $M^{\rm sat}(T)/M^{\rm sat}(5\,{\rm K})$.
		Both quantities decrease with increasing temperature nearly linearly and vanish near \Tc.
		These common temperature dependences provide an additional demonstration that the remanent Kerr response originates from vortex pinning.

		Figure~\ref{fig:4}(f) shows the temperature dependence of $H_{\rm p}$ extracted from the MOKE and magnetization data.
		Notice that, for the MOKE data, the fit gives the local scale $H_{\rm spot}(T)$ in the function $F_0(|H_{\rm T}|/H_{\rm spot})$.
		We convert this local scale to $H_{\rm p}$ using
		\begin{equation}
			H_{\rm p}^{\rm MOKE}(T)
			=
			\frac{L}{s_{\rm spot}}
			H_{\rm spot}(T)
			\simeq
			5H_{\rm spot}(T).
		\end{equation}
		Here, we use $L\simeq\SI{0.5}{\milli\meter}$, corresponding to approximately half of the lateral sample dimension, and $s_{\rm spot}\simeq\SI{0.1}{\milli\meter}$, the distance from the nearest sample edge to the optical spot [Fig.~\ref{fig:1}(b)].
		For the magnetization data, $H_{\rm p}^{M}$ is obtained directly as $H_{\rm p}$ entering the bulk-averaged function $F_2(|H_{\rm T}|/H_{\rm p})$.
		For both MOKE and magnetization measurements, the magnitudes of $H_{\rm p}^{\rm MOKE}$ and $H_{\rm p}^{M}$ fall within the same order of magnitude, although they do not coincide perfectly.
		Moreover, both $H_{\rm p}$ values decrease with increasing temperature and become small near \Tc.
		This level of agreement is reasonable because the analyses rely on a simplified Bean-model profile rather than on an exact critical-state solution for the actual sample geometry.

		The inset of Fig.~\ref{fig:4}(f) shows the magnetic-field dependence of the critical current density $j_{\rm c}(H)$ at \SI{5}{\kelvin}, estimated from the magnetization hysteresis loop.
		In SI units, for a rectangular sample, $j_{\rm c}(H)$ can be estimated as
		\begin{equation}
			j_{\rm c}(H)
			=
			\frac{2\Delta M}{a(1-a/3b)},
		\end{equation}
		where $\Delta M$ is the full width of the magnetization hysteresis loop expressed in units of \si{\ampere\per\meter}~\cite{Kim1963-xe,Chen1990-yo,Kim2000-sy,Tao2015-sm}.
		Our estimate of $j_{\rm c}(0)\sim\SI{8.4e8}{\ampere\per\meter\squared}$ at
		\SI{5}{\kelvin} is comparable to values reported in previous magnetization studies of BKBO:
		$\SI{1.5e9}{\ampere\per\meter\squared}$ at \SI{5}{\kelvin} in zero field for
		$x\sim0.37$~\cite{Kim2000-sy},
		$\SI{1e9}{\ampere\per\meter\squared}$ at \SI{2}{\kelvin} in zero field for
		$x\sim0.32$~\cite{Tao2015-sm},
		and $\SI{5.46e8}{\ampere\per\meter\squared}$ at \SI{4.2}{\kelvin} in zero field for
		$x\sim0.37$~\cite{Barilo1995-my}.

		Using the zero-field value of $j_{\rm c}\sim\SI{8.4e8}{\ampere\per\meter\squared}$ at \SI{5}{\kelvin}, we estimate the full critical-state penetration field as
		\begin{equation}
			\mu_0H_{\rm p}^{j_{\rm c}}
			\sim
			\mu_0j_{\rm c}(0)L= \SI{0.50}{\tesla}.
		\end{equation}
		This value is plotted in Fig.~\ref{fig:4}(f) with a star symbol.
		The experimental $H_{\rm p}$ values are comparable to this estimate, indicating the relevance of our analysis based on the Bean critical-state model.

	\subsection{Implications for MOKE studies to detect TRSB superconductivity}

		\begin{figure}[ht]
			\centering
			\includegraphics[width=246pt, bb=0.000000 0.000000 245.796874 124.147621]{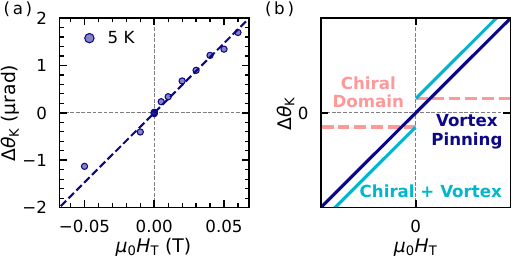}
			\caption{
				(a) Low training-field dependence of the remanent Kerr response $\Delta\theta_{\rm K}$ for BKBO at \SI{5}{\kelvin}.
				The response increases approximately linearly with the training field near zero field, without showing a pronounced anomalous enhancement expected for TRSB phenomena such as domain alignment in chiral superconductivity.
				(b) Schematic illustration of the training-field dependence of the remanent Kerr response expected for chiral-domain alignment, trapped vortices, and their coexistence.
				The chiral-domain alignment is expected to produce a pronounced step-like anomaly of the Kerr response at low training fields (dashed curve), whereas the trapped vortices are expected to produce a monotonic increase without any low-field anomalies (blue line).
				The absence of such a low-field enhancement in the present data is observed in BKBO.
			}\label{fig:5}
		\end{figure}

		Our results also have important implications for MOKE studies of TRSB superconductivity.
		In such experiments, a remanent Kerr signal is often discussed as possible evidence for a magnetic response associated with a TRSB superconducting state.
		Field training is often required to align chiral domains in TRSB superconductors~\cite{Xia2006-rk,Kapitulnik2009-xh,Schemm2014-cz,Schemm2015-tr}, because the MOKE response arising from the randomly oriented domains is expected to be significantly reduced if the domain size is much smaller than the optical spot size.
		However, the present results demonstrate that trapped vortices can also generate a sizable magnetic-history-dependent Kerr response after field training.
		Therefore, vortex pinning must be carefully distinguished from TRSB effects when interpreting remanent Kerr signals in type-II superconductors.

		Figure~\ref{fig:5}(a) focuses on the low-training-field region, which is important for the distinction.
		In BKBO, we find that the remanent Kerr response changes approximately linearly with $H_{\rm T}$ near zero $H_{\rm T}$ and shows no additional low-field anomalies.
		This behavior is naturally expected for trapped vortices, because the amount of trapped flux should increase continuously and linearly with the training field.
		In contrast, TRSB-domain alignment would produce a much sharper low-field anomaly if a weak training field biases the domain population (dashed curve in Fig.~\ref{fig:5}(b)).
		Indeed, such $H_{\rm T}$-independent step-like behavior has been reported in studies of Sr$_2$RuO$_4$~\cite{Xia2006-rk} and \ce{UPt3}~\cite{Schemm2014-cz}.
		If both effects coexist, the vortex contribution should provide a smooth background, while the TRSB-domain contribution would add an extra low-field step, as schematically shown by the cyan curve in Fig.~\ref{fig:5}(b).
		A similar issue has been discussed in \ce{UTe2}, where field-trained MOKE measurements have been interpreted in terms of both TRSB and vortex-induced critical-state effects~\cite{Wei2022-hn,Ajeesh2023-kz}.
		The absence of such an additional low-field feature in the present data shows that time-reversal symmetry is preserved in the superconducting state of BKBO.

		These considerations show that systematic training-field dependence provides a practical diagnostic for separating vortex-induced Kerr signals from intrinsic TRSB responses.
		Measurements in the low-field region are especially important, because this is where chiral-domain alignment and vortex trapping are expected to differ most clearly.
		Careful control of remnant fields, as performed in this study using the movable magnetic shield, is also essential, since unintended residual fields can introduce trapped vortices and mimic a history-dependent Kerr signal.
		Comparison with bulk magnetization measurements provides an additional check by testing whether the Kerr response follows the behavior expected for vortex pinning and critical-state physics.

\section{Conclusion}\label{sec:conclusion}
	In conclusion, we have investigated vortex pinning in the three-dimensional oxide superconductor \BKBOx{0.38}.
	We combined ultra-high-resolution MOKE measurements with detailed magnetization measurements.
	We found that the Kerr signal measured during ZFW after FC exhibits a clear magnetic-history dependence in the superconducting state.
	This remanent response disappears above \Tc.
	The overall behavior is qualitatively consistent with the remanent magnetization measured under the same field-training protocol.
	This correspondence indicates that the observed MOKE signal originates from the local remanent magnetic flux generated by trapped vortices.

	We further showed that the temperature and training-field dependences of the remanent MOKE response can be consistently understood using a Bean-model description of vortex pinning.
	In this picture, local MOKE sampling and bulk magnetization measurements are described by different spatial weighting of the same nonuniform remanent-flux profile.
	The characteristic full-penetration fields extracted from the MOKE and magnetization data are reasonably consistent with the Bean-model field scale estimated from the critical current density.
	These results demonstrate that MOKE provides a viable and sensitive optical probe of vortex pinning in type-II superconductors.
	Because MOKE probes a mesoscopic region without electrical contacts and can be performed rapidly, it may also provide a useful tool for evaluating vortex-pinning properties in superconducting materials for engineering applications.

	We also show that $\Delta\theta_{\rm K}$ of BKBO depends approximately linearly on the training field near zero field.
	This behavior is inconsistent with pronounced low-field anomalies expected for TRSB phenomena such as chiral-domain alignment.
	This provides an important basis for future MOKE studies of unconventional superconductors.

\section*{Acknowledgements}
	We acknowledge I.~Kakeya and Y.~Gotoh for valuable comments and discussions.
	We also thank Y. Hu, K.~Yada, J.~Xia, C.~Farhang, and J.~Wang for technical assistance in the early stage of the introduction of the MOKE technique to the Kyoto Univ. Group.

	This work was supported by Grant-in-Aids for Academic Transformation Area Research (A) ``Quantum Asymmetry'' (KAKENHI Grant Nos.~JP23H04866, JP23H04869, JP24H01663) from the Japan Society for the Promotion of Science (JSPS),
	by Grant-in-Aids for Scientific Research (KAKENHI Grant Nos.~JP23K17670, JP23H04861, JP24H00194, JP25K00960, JP25K00955, JP25H01246, JP25H01247) from JSPS,
	by ISHIZUE 2023 of Kyoto University Research Development Program,
	by Iketani Science and Technology Foundation (Grant No.~0361078-A),
	and by The Mitsubishi Foundation (Grant No.~202410051).
	The work at the University of Minnesota was funded by the US Department of Energy through the University of Minnesota Center for Quantum Materials, under Grant No.~DE-SC-0016371.
	The work at Columbia University was supported by the US NSF grant DMR-2104661.
	We acknowledge support for the construction of experimental setups from Research Equipment Development Support Room of the Graduate School of Science, Kyoto University; and support for liquid helium and nitrogen supplies from Low Temperature and Materials Sciences Division, Agency for Health, Safety and Environment, Kyoto University.

	\bibliographystyle{apsrev4-2-arxivdot}
	\bibliography{aps-control,paperpile}

\begin{thebibliography}{52}%
\makeatletter
\providecommand \@ifxundefined [1]{%
 \@ifx{#1\undefined}
}%
\providecommand \@ifnum [1]{%
 \ifnum #1\expandafter \@firstoftwo
 \else \expandafter \@secondoftwo
 \fi
}%
\providecommand \@ifx [1]{%
 \ifx #1\expandafter \@firstoftwo
 \else \expandafter \@secondoftwo
 \fi
}%
\providecommand \natexlab [1]{#1}%
\providecommand \enquote  [1]{``#1''}%
\providecommand \bibnamefont  [1]{#1}%
\providecommand \bibfnamefont [1]{#1}%
\providecommand \citenamefont [1]{#1}%
\providecommand \href@noop [0]{\@secondoftwo}%
\providecommand \href [0]{\begingroup \@sanitize@url \@href}%
\providecommand \@href[1]{\@@startlink{#1}\@@href}%
\providecommand \@@href[1]{\endgroup#1\@@endlink}%
\providecommand \@sanitize@url [0]{\catcode `\\12\catcode `\$12\catcode `\&12\catcode `\#12\catcode `\^12\catcode `\_12\catcode `\%12\relax}%
\providecommand \@@startlink[1]{}%
\providecommand \@@endlink[0]{}%
\providecommand \url  [0]{\begingroup\@sanitize@url \@url }%
\providecommand \@url [1]{\endgroup\@href {#1}{\urlprefix }}%
\providecommand \urlprefix  [0]{URL }%
\providecommand \Eprint [0]{\href }%
\providecommand \doibase [0]{https://doi.org/}%
\providecommand \selectlanguage [0]{\@gobble}%
\providecommand \bibinfo  [0]{\@secondoftwo}%
\providecommand \bibfield  [0]{\@secondoftwo}%
\providecommand \translation [1]{[#1]}%
\providecommand \BibitemOpen [0]{}%
\providecommand \bibitemStop [0]{}%
\providecommand \bibitemNoStop [0]{.\EOS\space}%
\providecommand \EOS [0]{\spacefactor3000\relax}%
\providecommand \BibitemShut  [1]{\csname bibitem#1\endcsname}%
\let\auto@bib@innerbib\@empty
\bibitem [{\citenamefont {Blatter}\ \emph {et~al.}(1994)\citenamefont {Blatter}, \citenamefont {Feigel'man}, \citenamefont {Geshkenbein}, \citenamefont {Larkin},\ and\ \citenamefont {Vinokur}}]{Blatter1994-ve}%
  \BibitemOpen
  \bibfield  {author} {\bibinfo {author} {\bibfnamefont {G.}~\bibnamefont {Blatter}}, \bibinfo {author} {\bibfnamefont {M.~V.}\ \bibnamefont {Feigel'man}}, \bibinfo {author} {\bibfnamefont {V.~B.}\ \bibnamefont {Geshkenbein}}, \bibinfo {author} {\bibfnamefont {A.~I.}\ \bibnamefont {Larkin}},\ and\ \bibinfo {author} {\bibfnamefont {V.~M.}\ \bibnamefont {Vinokur}},\ }\href {https://doi.org/10.1103/revmodphys.66.1125} {\bibfield  {journal} {\bibinfo  {journal} {Rev. Mod. Phys.}\ }\textbf {\bibinfo {volume} {66}},\ \bibinfo {pages} {1125} (\bibinfo {year} {1994})}\BibitemShut {NoStop}%
\bibitem [{\citenamefont {Bean}(1964)}]{Bean1964-bt}%
  \BibitemOpen
  \bibfield  {author} {\bibinfo {author} {\bibfnamefont {C.~P.}\ \bibnamefont {Bean}},\ }\href {https://doi.org/10.1103/revmodphys.36.31} {\bibfield  {journal} {\bibinfo  {journal} {Rev. Mod. Phys.}\ }\textbf {\bibinfo {volume} {36}},\ \bibinfo {pages} {31} (\bibinfo {year} {1964})}\BibitemShut {NoStop}%
\bibitem [{\citenamefont {Eley}\ \emph {et~al.}(2021)\citenamefont {Eley}, \citenamefont {Glatz},\ and\ \citenamefont {Willa}}]{Eley2021-zm}%
  \BibitemOpen
  \bibfield  {author} {\bibinfo {author} {\bibfnamefont {S.}~\bibnamefont {Eley}}, \bibinfo {author} {\bibfnamefont {A.}~\bibnamefont {Glatz}},\ and\ \bibinfo {author} {\bibfnamefont {R.}~\bibnamefont {Willa}},\ }\href {https://doi.org/10.1063/5.0055611} {\bibfield  {journal} {\bibinfo  {journal} {J. Appl. Phys.}\ }\textbf {\bibinfo {volume} {130}},\ \bibinfo {pages} {050901} (\bibinfo {year} {2021})}\BibitemShut {NoStop}%
\bibitem [{\citenamefont {Kwok}\ \emph {et~al.}(2016)\citenamefont {Kwok}, \citenamefont {Welp}, \citenamefont {Glatz}, \citenamefont {Koshelev}, \citenamefont {Kihlstrom},\ and\ \citenamefont {Crabtree}}]{Kwok2016-lc}%
  \BibitemOpen
  \bibfield  {author} {\bibinfo {author} {\bibfnamefont {W.-K.}\ \bibnamefont {Kwok}}, \bibinfo {author} {\bibfnamefont {U.}~\bibnamefont {Welp}}, \bibinfo {author} {\bibfnamefont {A.}~\bibnamefont {Glatz}}, \bibinfo {author} {\bibfnamefont {A.~E.}\ \bibnamefont {Koshelev}}, \bibinfo {author} {\bibfnamefont {K.~J.}\ \bibnamefont {Kihlstrom}},\ and\ \bibinfo {author} {\bibfnamefont {G.~W.}\ \bibnamefont {Crabtree}},\ }\href {https://doi.org/10.1088/0034-4885/79/11/116501} {\bibfield  {journal} {\bibinfo  {journal} {Rep. Prog. Phys.}\ }\textbf {\bibinfo {volume} {79}},\ \bibinfo {pages} {116501} (\bibinfo {year} {2016})}\BibitemShut {NoStop}%
\bibitem [{\citenamefont {Sigrist}\ and\ \citenamefont {Ueda}(1991)}]{Sigrist1991-nf}%
  \BibitemOpen
  \bibfield  {author} {\bibinfo {author} {\bibfnamefont {M.}~\bibnamefont {Sigrist}}\ and\ \bibinfo {author} {\bibfnamefont {K.}~\bibnamefont {Ueda}},\ }\href {https://doi.org/10.1103/revmodphys.63.239} {\bibfield  {journal} {\bibinfo  {journal} {Rev. Mod. Phys.}\ }\textbf {\bibinfo {volume} {63}},\ \bibinfo {pages} {239} (\bibinfo {year} {1991})}\BibitemShut {NoStop}%
\bibitem [{\citenamefont {Kallin}\ and\ \citenamefont {Berlinsky}(2016)}]{Kallin2016-wg}%
  \BibitemOpen
  \bibfield  {author} {\bibinfo {author} {\bibfnamefont {C.}~\bibnamefont {Kallin}}\ and\ \bibinfo {author} {\bibfnamefont {J.}~\bibnamefont {Berlinsky}},\ }\href {https://doi.org/10.1088/0034-4885/79/5/054502} {\bibfield  {journal} {\bibinfo  {journal} {Rep. Prog. Phys.}\ }\textbf {\bibinfo {volume} {79}},\ \bibinfo {pages} {054502} (\bibinfo {year} {2016})}\BibitemShut {NoStop}%
\bibitem [{\citenamefont {Ghosh}\ \emph {et~al.}(2020)\citenamefont {Ghosh}, \citenamefont {Smidman}, \citenamefont {Shang}, \citenamefont {Annett}, \citenamefont {Hillier}, \citenamefont {Quintanilla},\ and\ \citenamefont {Yuan}}]{Ghosh2020-ky}%
  \BibitemOpen
  \bibfield  {author} {\bibinfo {author} {\bibfnamefont {S.~K.}\ \bibnamefont {Ghosh}}, \bibinfo {author} {\bibfnamefont {M.}~\bibnamefont {Smidman}}, \bibinfo {author} {\bibfnamefont {T.}~\bibnamefont {Shang}}, \bibinfo {author} {\bibfnamefont {J.~F.}\ \bibnamefont {Annett}}, \bibinfo {author} {\bibfnamefont {A.~D.}\ \bibnamefont {Hillier}}, \bibinfo {author} {\bibfnamefont {J.}~\bibnamefont {Quintanilla}},\ and\ \bibinfo {author} {\bibfnamefont {H.}~\bibnamefont {Yuan}},\ }\href {https://doi.org/10.1088/1361-648X/abaa06} {\bibfield  {journal} {\bibinfo  {journal} {J. Phys. Condens. Matter}\ }\textbf {\bibinfo {volume} {33}},\ \bibinfo {pages} {033001} (\bibinfo {year} {2020})}\BibitemShut {NoStop}%
\bibitem [{\citenamefont {Cava}\ \emph {et~al.}(1988)\citenamefont {Cava}, \citenamefont {Batlogg}, \citenamefont {Krajewski}, \citenamefont {Farrow}, \citenamefont {Rupp}, \citenamefont {White}, \citenamefont {Short}, \citenamefont {Peck},\ and\ \citenamefont {Kometani}}]{Cava1988-vn}%
  \BibitemOpen
  \bibfield  {author} {\bibinfo {author} {\bibfnamefont {R.~J.}\ \bibnamefont {Cava}}, \bibinfo {author} {\bibfnamefont {B.}~\bibnamefont {Batlogg}}, \bibinfo {author} {\bibfnamefont {J.~J.}\ \bibnamefont {Krajewski}}, \bibinfo {author} {\bibfnamefont {R.}~\bibnamefont {Farrow}}, \bibinfo {author} {\bibfnamefont {L.~W.}\ \bibnamefont {Rupp}, \bibfnamefont {Jr}}, \bibinfo {author} {\bibfnamefont {A.~E.}\ \bibnamefont {White}}, \bibinfo {author} {\bibfnamefont {K.}~\bibnamefont {Short}}, \bibinfo {author} {\bibfnamefont {W.~F.}\ \bibnamefont {Peck}},\ and\ \bibinfo {author} {\bibfnamefont {T.}~\bibnamefont {Kometani}},\ }\href {https://doi.org/10.1038/332814a0} {\bibfield  {journal} {\bibinfo  {journal} {Nature}\ }\textbf {\bibinfo {volume} {332}},\ \bibinfo {pages} {814} (\bibinfo {year} {1988})}\BibitemShut {NoStop}%
\bibitem [{\citenamefont {Hinks}\ \emph {et~al.}(1988)\citenamefont {Hinks}, \citenamefont {Dabrowski}, \citenamefont {Jorgensen}, \citenamefont {Mitchell}, \citenamefont {Richards}, \citenamefont {Pei},\ and\ \citenamefont {Shi}}]{Hinks1988-au}%
  \BibitemOpen
  \bibfield  {author} {\bibinfo {author} {\bibfnamefont {D.~G.}\ \bibnamefont {Hinks}}, \bibinfo {author} {\bibfnamefont {B.}~\bibnamefont {Dabrowski}}, \bibinfo {author} {\bibfnamefont {J.~D.}\ \bibnamefont {Jorgensen}}, \bibinfo {author} {\bibfnamefont {A.~W.}\ \bibnamefont {Mitchell}}, \bibinfo {author} {\bibfnamefont {D.~R.}\ \bibnamefont {Richards}}, \bibinfo {author} {\bibfnamefont {S.}~\bibnamefont {Pei}},\ and\ \bibinfo {author} {\bibfnamefont {D.}~\bibnamefont {Shi}},\ }\href {https://doi.org/10.1038/333836a0} {\bibfield  {journal} {\bibinfo  {journal} {Nature}\ }\textbf {\bibinfo {volume} {333}},\ \bibinfo {pages} {836} (\bibinfo {year} {1988})}\BibitemShut {NoStop}%
\bibitem [{\citenamefont {Sleight}(2015)}]{Sleight2015-wr}%
  \BibitemOpen
  \bibfield  {author} {\bibinfo {author} {\bibfnamefont {A.~W.}\ \bibnamefont {Sleight}},\ }\href {https://doi.org/10.1016/j.physc.2015.02.012} {\bibfield  {journal} {\bibinfo  {journal} {Physica C Supercond.}\ }\textbf {\bibinfo {volume} {514}},\ \bibinfo {pages} {152} (\bibinfo {year} {2015})}\BibitemShut {NoStop}%
\bibitem [{\citenamefont {Griffitt}\ \emph {et~al.}(2023)\citenamefont {Griffitt}, \citenamefont {Spai\'{c}}, \citenamefont {Joe}, \citenamefont {Anderson}, \citenamefont {Zhai}, \citenamefont {Krogstad}, \citenamefont {Osborn}, \citenamefont {Pelc},\ and\ \citenamefont {Greven}}]{Griffitt2023-cp}%
  \BibitemOpen
  \bibfield  {author} {\bibinfo {author} {\bibfnamefont {S.}~\bibnamefont {Griffitt}}, \bibinfo {author} {\bibfnamefont {M.}~\bibnamefont {Spai\'{c}}}, \bibinfo {author} {\bibfnamefont {J.}~\bibnamefont {Joe}}, \bibinfo {author} {\bibfnamefont {Z.~W.}\ \bibnamefont {Anderson}}, \bibinfo {author} {\bibfnamefont {D.}~\bibnamefont {Zhai}}, \bibinfo {author} {\bibfnamefont {M.~J.}\ \bibnamefont {Krogstad}}, \bibinfo {author} {\bibfnamefont {R.}~\bibnamefont {Osborn}}, \bibinfo {author} {\bibfnamefont {D.}~\bibnamefont {Pelc}},\ and\ \bibinfo {author} {\bibfnamefont {M.}~\bibnamefont {Greven}},\ }\href {https://doi.org/10.1038/s41467-023-36348-9} {\bibfield  {journal} {\bibinfo  {journal} {Nat. Commun.}\ }\textbf {\bibinfo {volume} {14}},\ \bibinfo {pages} {845} (\bibinfo {year} {2023})}\BibitemShut {NoStop}%
\bibitem [{\citenamefont {Barilo}\ \emph {et~al.}(1999)\citenamefont {Barilo}, \citenamefont {Shiryaev}, \citenamefont {Gatalskaya}, \citenamefont {Zhigunov}, \citenamefont {Pushkarev}, \citenamefont {Fedotova}, \citenamefont {Szymczak}, \citenamefont {Szymczak}, \citenamefont {Baran}, \citenamefont {Lynn}, \citenamefont {Rosov},\ and\ \citenamefont {Skanthakumar}}]{Barilo1999-gl}%
  \BibitemOpen
  \bibfield  {author} {\bibinfo {author} {\bibfnamefont {S.~N.}\ \bibnamefont {Barilo}}, \bibinfo {author} {\bibfnamefont {S.~V.}\ \bibnamefont {Shiryaev}}, \bibinfo {author} {\bibfnamefont {V.~I.}\ \bibnamefont {Gatalskaya}}, \bibinfo {author} {\bibfnamefont {D.~I.}\ \bibnamefont {Zhigunov}}, \bibinfo {author} {\bibfnamefont {A.~V.}\ \bibnamefont {Pushkarev}}, \bibinfo {author} {\bibfnamefont {V.~V.}\ \bibnamefont {Fedotova}}, \bibinfo {author} {\bibfnamefont {H.}~\bibnamefont {Szymczak}}, \bibinfo {author} {\bibfnamefont {R.}~\bibnamefont {Szymczak}}, \bibinfo {author} {\bibfnamefont {M.}~\bibnamefont {Baran}}, \bibinfo {author} {\bibfnamefont {J.~W.}\ \bibnamefont {Lynn}}, \bibinfo {author} {\bibfnamefont {N.}~\bibnamefont {Rosov}},\ and\ \bibinfo {author} {\bibfnamefont {S.}~\bibnamefont {Skanthakumar}},\ }\href {https://doi.org/10.1016/s0022-0248(98)01201-9} {\bibfield  {journal} {\bibinfo  {journal} {J. Cryst. Growth}\ }\textbf {\bibinfo {volume} {198-199}},\ \bibinfo {pages} {636} (\bibinfo {year} {1999})}\BibitemShut {NoStop}%
\bibitem [{\citenamefont {Kumar}\ \emph {et~al.}(1999)\citenamefont {Kumar}, \citenamefont {Hall},\ and\ \citenamefont {Goodrich}}]{Kumar1999-gz}%
  \BibitemOpen
  \bibfield  {author} {\bibinfo {author} {\bibfnamefont {P.}~\bibnamefont {Kumar}}, \bibinfo {author} {\bibfnamefont {D.}~\bibnamefont {Hall}},\ and\ \bibinfo {author} {\bibfnamefont {R.~G.}\ \bibnamefont {Goodrich}},\ }\href {https://doi.org/10.1103/physrevlett.82.4532} {\bibfield  {journal} {\bibinfo  {journal} {Phys. Rev. Lett.}\ }\textbf {\bibinfo {volume} {82}},\ \bibinfo {pages} {4532} (\bibinfo {year} {1999})}\BibitemShut {NoStop}%
\bibitem [{\citenamefont {Mosley}\ \emph {et~al.}(1994)\citenamefont {Mosley}, \citenamefont {Dykes}, \citenamefont {Shelton}, \citenamefont {Sterne},\ and\ \citenamefont {Howell}}]{Mosley1994-rj}%
  \BibitemOpen
  \bibfield  {author} {\bibinfo {author} {\bibfnamefont {W.~D.}\ \bibnamefont {Mosley}}, \bibinfo {author} {\bibfnamefont {J.~W.}\ \bibnamefont {Dykes}}, \bibinfo {author} {\bibfnamefont {R.~N.}\ \bibnamefont {Shelton}}, \bibinfo {author} {\bibfnamefont {A.}~\bibnamefont {Sterne}},\ and\ \bibinfo {author} {\bibfnamefont {R.~H.}\ \bibnamefont {Howell}},\ }\href {https://doi.org/10.1103/PhysRevLett.73.1271} {\bibfield  {journal} {\bibinfo  {journal} {Phys. Rev. Lett.}\ }\textbf {\bibinfo {volume} {73}},\ \bibinfo {pages} {1271} (\bibinfo {year} {1994})}\BibitemShut {NoStop}%
\bibitem [{\citenamefont {Wen}\ \emph {et~al.}(2018)\citenamefont {Wen}, \citenamefont {Xu}, \citenamefont {Yao}, \citenamefont {Peng}, \citenamefont {Niu}, \citenamefont {Chen}, \citenamefont {Liu}, \citenamefont {Shen}, \citenamefont {Song}, \citenamefont {Lou}, \citenamefont {Fang}, \citenamefont {Liu}, \citenamefont {Song}, \citenamefont {Jiao}, \citenamefont {Duan}, \citenamefont {Wen}, \citenamefont {Dudin}, \citenamefont {Kotliar}, \citenamefont {Yin},\ and\ \citenamefont {Feng}}]{Wen2018-wi}%
  \BibitemOpen
  \bibfield  {author} {\bibinfo {author} {\bibfnamefont {C.~H.~P.}\ \bibnamefont {Wen}}, \bibinfo {author} {\bibfnamefont {H.~C.}\ \bibnamefont {Xu}}, \bibinfo {author} {\bibfnamefont {Q.}~\bibnamefont {Yao}}, \bibinfo {author} {\bibfnamefont {R.}~\bibnamefont {Peng}}, \bibinfo {author} {\bibfnamefont {X.~H.}\ \bibnamefont {Niu}}, \bibinfo {author} {\bibfnamefont {Q.~Y.}\ \bibnamefont {Chen}}, \bibinfo {author} {\bibfnamefont {Z.~T.}\ \bibnamefont {Liu}}, \bibinfo {author} {\bibfnamefont {D.~W.}\ \bibnamefont {Shen}}, \bibinfo {author} {\bibfnamefont {Q.}~\bibnamefont {Song}}, \bibinfo {author} {\bibfnamefont {X.}~\bibnamefont {Lou}}, \bibinfo {author} {\bibfnamefont {Y.~F.}\ \bibnamefont {Fang}}, \bibinfo {author} {\bibfnamefont {X.~S.}\ \bibnamefont {Liu}}, \bibinfo {author} {\bibfnamefont {Y.~H.}\ \bibnamefont {Song}}, \bibinfo {author} {\bibfnamefont {Y.~J.}\ \bibnamefont {Jiao}}, \bibinfo {author} {\bibfnamefont {T.~F.}\ \bibnamefont {Duan}}, \bibinfo {author} {\bibfnamefont {H.~H.}\ \bibnamefont {Wen}}, \bibinfo {author} {\bibfnamefont {P.}~\bibnamefont {Dudin}}, \bibinfo {author} {\bibfnamefont {G.}~\bibnamefont {Kotliar}}, \bibinfo {author} {\bibfnamefont {Z.~P.}\ \bibnamefont {Yin}},\ and\ \bibinfo {author} {\bibfnamefont {D.~L.}\ \bibnamefont {Feng}},\ }\href {https://doi.org/10.1103/PhysRevLett.121.117002} {\bibfield  {journal} {\bibinfo  {journal} {Phys. Rev. Lett.}\ }\textbf {\bibinfo {volume} {121}},\ \bibinfo {pages} {117002} (\bibinfo {year} {2018})}\BibitemShut {NoStop}%
\bibitem [{\citenamefont {Bhattacharyya}\ \emph {et~al.}(2025)\citenamefont {Bhattacharyya}, \citenamefont {Thangavel},\ and\ \citenamefont {Sarun}}]{Bhattacharyya2025-if}%
  \BibitemOpen
  \bibfield  {author} {\bibinfo {author} {\bibfnamefont {S.}~\bibnamefont {Bhattacharyya}}, \bibinfo {author} {\bibfnamefont {R.}~\bibnamefont {Thangavel}},\ and\ \bibinfo {author} {\bibfnamefont {P.~M.}\ \bibnamefont {Sarun}},\ }\href {https://doi.org/10.1063/5.0260314} {\bibfield  {journal} {\bibinfo  {journal} {J. Appl. Phys.}\ }\textbf {\bibinfo {volume} {137}},\ \bibinfo {pages} {155902} (\bibinfo {year} {2025})}\BibitemShut {NoStop}%
\bibitem [{\citenamefont {Damascelli}\ \emph {et~al.}(2003)\citenamefont {Damascelli}, \citenamefont {Hussain},\ and\ \citenamefont {Shen}}]{Damascelli2003-cw}%
  \BibitemOpen
  \bibfield  {author} {\bibinfo {author} {\bibfnamefont {A.}~\bibnamefont {Damascelli}}, \bibinfo {author} {\bibfnamefont {Z.}~\bibnamefont {Hussain}},\ and\ \bibinfo {author} {\bibfnamefont {Z.-X.}\ \bibnamefont {Shen}},\ }\href {https://doi.org/10.1103/revmodphys.75.473} {\bibfield  {journal} {\bibinfo  {journal} {Rev. Mod. Phys.}\ }\textbf {\bibinfo {volume} {75}},\ \bibinfo {pages} {473} (\bibinfo {year} {2003})}\BibitemShut {NoStop}%
\bibitem [{\citenamefont {Shibauchi}\ \emph {et~al.}(2020)\citenamefont {Shibauchi}, \citenamefont {Hanaguri},\ and\ \citenamefont {Matsuda}}]{Shibauchi2020-hf}%
  \BibitemOpen
  \bibfield  {author} {\bibinfo {author} {\bibfnamefont {T.}~\bibnamefont {Shibauchi}}, \bibinfo {author} {\bibfnamefont {T.}~\bibnamefont {Hanaguri}},\ and\ \bibinfo {author} {\bibfnamefont {Y.}~\bibnamefont {Matsuda}},\ }\href {https://doi.org/10.7566/JPSJ.89.102002} {\bibfield  {journal} {\bibinfo  {journal} {J. Phys. Soc. Jpn.}\ }\textbf {\bibinfo {volume} {89}},\ \bibinfo {pages} {102002} (\bibinfo {year} {2020})}\BibitemShut {NoStop}%
\bibitem [{\citenamefont {Hirschfeld}\ \emph {et~al.}(2011)\citenamefont {Hirschfeld}, \citenamefont {Korshunov},\ and\ \citenamefont {Mazin}}]{Hirschfeld2011-iv}%
  \BibitemOpen
  \bibfield  {author} {\bibinfo {author} {\bibfnamefont {P.~J.}\ \bibnamefont {Hirschfeld}}, \bibinfo {author} {\bibfnamefont {M.~M.}\ \bibnamefont {Korshunov}},\ and\ \bibinfo {author} {\bibfnamefont {I.~I.}\ \bibnamefont {Mazin}},\ }\href {https://doi.org/10.1088/0034-4885/74/12/124508} {\bibfield  {journal} {\bibinfo  {journal} {Rep. Prog. Phys.}\ }\textbf {\bibinfo {volume} {74}},\ \bibinfo {pages} {124508} (\bibinfo {year} {2011})}\BibitemShut {NoStop}%
\bibitem [{\citenamefont {Hess}\ \emph {et~al.}(1989)\citenamefont {Hess}, \citenamefont {Robinson}, \citenamefont {Dynes}, \citenamefont {Valles},\ and\ \citenamefont {Waszczak}}]{Hess1989-zw}%
  \BibitemOpen
  \bibfield  {author} {\bibinfo {author} {\bibfnamefont {H.~F.}\ \bibnamefont {Hess}}, \bibinfo {author} {\bibfnamefont {R.~B.}\ \bibnamefont {Robinson}}, \bibinfo {author} {\bibfnamefont {R.~C.}\ \bibnamefont {Dynes}}, \bibinfo {author} {\bibfnamefont {J.~M.}\ \bibnamefont {Valles}, \bibfnamefont {Jr}},\ and\ \bibinfo {author} {\bibfnamefont {J.~V.}\ \bibnamefont {Waszczak}},\ }\href {https://doi.org/10.1103/PhysRevLett.62.214} {\bibfield  {journal} {\bibinfo  {journal} {Phys. Rev. Lett.}\ }\textbf {\bibinfo {volume} {62}},\ \bibinfo {pages} {214} (\bibinfo {year} {1989})}\BibitemShut {NoStop}%
\bibitem [{\citenamefont {Kirtley}(2010)}]{Kirtley2010-ow}%
  \BibitemOpen
  \bibfield  {author} {\bibinfo {author} {\bibfnamefont {J.~R.}\ \bibnamefont {Kirtley}},\ }\href {https://doi.org/10.1088/0034-4885/73/12/126501} {\bibfield  {journal} {\bibinfo  {journal} {Rep. Prog. Phys.}\ }\textbf {\bibinfo {volume} {73}},\ \bibinfo {pages} {126501} (\bibinfo {year} {2010})}\BibitemShut {NoStop}%
\bibitem [{\citenamefont {Jooss}\ \emph {et~al.}(2002)\citenamefont {Jooss}, \citenamefont {Albrecht}, \citenamefont {Kuhn}, \citenamefont {Leonhardt},\ and\ \citenamefont {Kronm{\"{u}}ller}}]{Jooss2002-oj}%
  \BibitemOpen
  \bibfield  {author} {\bibinfo {author} {\bibfnamefont {C.}~\bibnamefont {Jooss}}, \bibinfo {author} {\bibfnamefont {J.}~\bibnamefont {Albrecht}}, \bibinfo {author} {\bibfnamefont {H.}~\bibnamefont {Kuhn}}, \bibinfo {author} {\bibfnamefont {S.}~\bibnamefont {Leonhardt}},\ and\ \bibinfo {author} {\bibfnamefont {H.}~\bibnamefont {Kronm{\"{u}}ller}},\ }\href {https://doi.org/10.1088/0034-4885/65/5/202} {\bibfield  {journal} {\bibinfo  {journal} {Rep. Prog. Phys.}\ }\textbf {\bibinfo {volume} {65}},\ \bibinfo {pages} {651} (\bibinfo {year} {2002})}\BibitemShut {NoStop}%
\bibitem [{\citenamefont {Goa}\ \emph {et~al.}(2001)\citenamefont {Goa}, \citenamefont {Hauglin}, \citenamefont {Baziljevich}, \citenamefont {Il'yashenko}, \citenamefont {Gammel},\ and\ \citenamefont {Johansen}}]{Goa2001-na}%
  \BibitemOpen
  \bibfield  {author} {\bibinfo {author} {\bibfnamefont {P.~E.}\ \bibnamefont {Goa}}, \bibinfo {author} {\bibfnamefont {H.}~\bibnamefont {Hauglin}}, \bibinfo {author} {\bibfnamefont {M.}~\bibnamefont {Baziljevich}}, \bibinfo {author} {\bibfnamefont {E.}~\bibnamefont {Il'yashenko}}, \bibinfo {author} {\bibfnamefont {P.~L.}\ \bibnamefont {Gammel}},\ and\ \bibinfo {author} {\bibfnamefont {T.~H.}\ \bibnamefont {Johansen}},\ }\href {https://doi.org/10.1088/0953-2048/14/9/320} {\bibfield  {journal} {\bibinfo  {journal} {Supercond. Sci. Technol.}\ }\textbf {\bibinfo {volume} {14}},\ \bibinfo {pages} {729} (\bibinfo {year} {2001})}\BibitemShut {NoStop}%
\bibitem [{\citenamefont {Kerr}(1877)}]{Kerr1877-tj}%
  \BibitemOpen
  \bibfield  {author} {\bibinfo {author} {\bibfnamefont {J.}~\bibnamefont {Kerr}},\ }\href {https://doi.org/10.1080/14786447708639245} {\bibfield  {journal} {\bibinfo  {journal} {Lond. Edinb. Dublin Philos. Mag. J. Sci.}\ }\textbf {\bibinfo {volume} {3}},\ \bibinfo {pages} {321} (\bibinfo {year} {1877})}\BibitemShut {NoStop}%
\bibitem [{\citenamefont {Qiu}\ and\ \citenamefont {Bader}(2000)}]{Qiu2000-wi}%
  \BibitemOpen
  \bibfield  {author} {\bibinfo {author} {\bibfnamefont {Z.~Q.}\ \bibnamefont {Qiu}}\ and\ \bibinfo {author} {\bibfnamefont {S.~D.}\ \bibnamefont {Bader}},\ }\href {https://doi.org/10.1063/1.1150496} {\bibfield  {journal} {\bibinfo  {journal} {Rev. Sci. Instrum.}\ }\textbf {\bibinfo {volume} {71}},\ \bibinfo {pages} {1243} (\bibinfo {year} {2000})}\BibitemShut {NoStop}%
\bibitem [{\citenamefont {Higo}\ \emph {et~al.}(2018)\citenamefont {Higo}, \citenamefont {Man}, \citenamefont {Gopman}, \citenamefont {Wu}, \citenamefont {Koretsune}, \citenamefont {van~'t Erve}, \citenamefont {Kabanov}, \citenamefont {Rees}, \citenamefont {Li}, \citenamefont {Suzuki}, \citenamefont {Patankar}, \citenamefont {Ikhlas}, \citenamefont {Chien}, \citenamefont {Arita}, \citenamefont {Shull}, \citenamefont {Orenstein},\ and\ \citenamefont {Nakatsuji}}]{Higo2018-yn}%
  \BibitemOpen
  \bibfield  {author} {\bibinfo {author} {\bibfnamefont {T.}~\bibnamefont {Higo}}, \bibinfo {author} {\bibfnamefont {H.}~\bibnamefont {Man}}, \bibinfo {author} {\bibfnamefont {D.~B.}\ \bibnamefont {Gopman}}, \bibinfo {author} {\bibfnamefont {L.}~\bibnamefont {Wu}}, \bibinfo {author} {\bibfnamefont {T.}~\bibnamefont {Koretsune}}, \bibinfo {author} {\bibfnamefont {O.~M.~J.}\ \bibnamefont {van~'t Erve}}, \bibinfo {author} {\bibfnamefont {Y.~P.}\ \bibnamefont {Kabanov}}, \bibinfo {author} {\bibfnamefont {D.}~\bibnamefont {Rees}}, \bibinfo {author} {\bibfnamefont {Y.}~\bibnamefont {Li}}, \bibinfo {author} {\bibfnamefont {M.-T.}\ \bibnamefont {Suzuki}}, \bibinfo {author} {\bibfnamefont {S.}~\bibnamefont {Patankar}}, \bibinfo {author} {\bibfnamefont {M.}~\bibnamefont {Ikhlas}}, \bibinfo {author} {\bibfnamefont {C.~L.}\ \bibnamefont {Chien}}, \bibinfo {author} {\bibfnamefont {R.}~\bibnamefont {Arita}}, \bibinfo {author} {\bibfnamefont {R.~D.}\ \bibnamefont {Shull}}, \bibinfo {author} {\bibfnamefont {J.}~\bibnamefont {Orenstein}},\ and\ \bibinfo {author} {\bibfnamefont {S.}~\bibnamefont {Nakatsuji}},\ }\href {https://doi.org/10.1038/s41566-017-0086-z} {\bibfield  {journal} {\bibinfo  {journal} {Nat. Photonics}\ }\textbf {\bibinfo {volume} {12}},\ \bibinfo {pages} {73} (\bibinfo {year} {2018})}\BibitemShut {NoStop}%
\bibitem [{\citenamefont {Kato}\ \emph {et~al.}(2023)\citenamefont {Kato}, \citenamefont {Okamura}, \citenamefont {Hirschberger}, \citenamefont {Tokura},\ and\ \citenamefont {Takahashi}}]{Kato2023-fp}%
  \BibitemOpen
  \bibfield  {author} {\bibinfo {author} {\bibfnamefont {Y.~D.}\ \bibnamefont {Kato}}, \bibinfo {author} {\bibfnamefont {Y.}~\bibnamefont {Okamura}}, \bibinfo {author} {\bibfnamefont {M.}~\bibnamefont {Hirschberger}}, \bibinfo {author} {\bibfnamefont {Y.}~\bibnamefont {Tokura}},\ and\ \bibinfo {author} {\bibfnamefont {Y.}~\bibnamefont {Takahashi}},\ }\href {https://doi.org/10.1038/s41467-023-41203-y} {\bibfield  {journal} {\bibinfo  {journal} {Nat. Commun.}\ }\textbf {\bibinfo {volume} {14}},\ \bibinfo {pages} {5416} (\bibinfo {year} {2023})}\BibitemShut {NoStop}%
\bibitem [{\citenamefont {Watanabe}\ \emph {et~al.}(2026)\citenamefont {Watanabe}, \citenamefont {Yamane}, \citenamefont {Maki}, \citenamefont {Ikeda}, \citenamefont {Kirikoshi}, \citenamefont {Otsuki}, \citenamefont {Aoyama}, \citenamefont {Ohgushi},\ and\ \citenamefont {Yonezawa}}]{Watanabe2026-sb}%
  \BibitemOpen
  \bibfield  {author} {\bibinfo {author} {\bibfnamefont {G.}~\bibnamefont {Watanabe}}, \bibinfo {author} {\bibfnamefont {S.}~\bibnamefont {Yamane}}, \bibinfo {author} {\bibfnamefont {R.}~\bibnamefont {Maki}}, \bibinfo {author} {\bibfnamefont {A.}~\bibnamefont {Ikeda}}, \bibinfo {author} {\bibfnamefont {A.}~\bibnamefont {Kirikoshi}}, \bibinfo {author} {\bibfnamefont {J.}~\bibnamefont {Otsuki}}, \bibinfo {author} {\bibfnamefont {T.}~\bibnamefont {Aoyama}}, \bibinfo {author} {\bibfnamefont {K.}~\bibnamefont {Ohgushi}},\ and\ \bibinfo {author} {\bibfnamefont {S.}~\bibnamefont {Yonezawa}}}, \href {https://arxiv.org/abs/2604.14947} {arXiv.2604.14947 (\bibinfo {year} {2026})}\BibitemShut {NoStop}%
\bibitem [{\citenamefont {Yang}\ \emph {et~al.}(2026)\citenamefont {Yang}, \citenamefont {Won}, \citenamefont {Cress}, \citenamefont {Franklin}, \citenamefont {Fang}, \citenamefont {Fields}, \citenamefont {Combs}, \citenamefont {Han}, \citenamefont {Lu}, \citenamefont {Bennett}, \citenamefont {Cheong},\ and\ \citenamefont {Xia}}]{Yang2026-lc}%
  \BibitemOpen
  \bibfield  {author} {\bibinfo {author} {\bibfnamefont {W.}~\bibnamefont {Yang}}, \bibinfo {author} {\bibfnamefont {C.}~\bibnamefont {Won}}, \bibinfo {author} {\bibfnamefont {C.}~\bibnamefont {Cress}}, \bibinfo {author} {\bibfnamefont {M.~Z.}\ \bibnamefont {Franklin}}, \bibinfo {author} {\bibfnamefont {X.}~\bibnamefont {Fang}}, \bibinfo {author} {\bibfnamefont {S.}~\bibnamefont {Fields}}, \bibinfo {author} {\bibfnamefont {N.}~\bibnamefont {Combs}}, \bibinfo {author} {\bibfnamefont {S.}~\bibnamefont {Han}}, \bibinfo {author} {\bibfnamefont {W.}~\bibnamefont {Lu}}, \bibinfo {author} {\bibfnamefont {S.~P.}\ \bibnamefont {Bennett}}, \bibinfo {author} {\bibfnamefont {S.-W.}\ \bibnamefont {Cheong}},\ and\ \bibinfo {author} {\bibfnamefont {J.}~\bibnamefont {Xia}}}, \href {https://arxiv.org/abs/2604.21021} {arXiv.2604.21021 (\bibinfo {year} {2026})}\BibitemShut {NoStop}%
\bibitem [{\citenamefont {Kapitulnik}\ \emph {et~al.}(2009)\citenamefont {Kapitulnik}, \citenamefont {Xia}, \citenamefont {Schemm},\ and\ \citenamefont {Palevski}}]{Kapitulnik2009-xh}%
  \BibitemOpen
  \bibfield  {author} {\bibinfo {author} {\bibfnamefont {A.}~\bibnamefont {Kapitulnik}}, \bibinfo {author} {\bibfnamefont {J.}~\bibnamefont {Xia}}, \bibinfo {author} {\bibfnamefont {E.}~\bibnamefont {Schemm}},\ and\ \bibinfo {author} {\bibfnamefont {A.}~\bibnamefont {Palevski}},\ }\href {https://doi.org/10.1088/1367-2630/11/5/055060} {\bibfield  {journal} {\bibinfo  {journal} {New J. Phys.}\ }\textbf {\bibinfo {volume} {11}},\ \bibinfo {pages} {055060} (\bibinfo {year} {2009})}\BibitemShut {NoStop}%
\bibitem [{\citenamefont {Xia}\ \emph {et~al.}(2006{\natexlab{a}})\citenamefont {Xia}, \citenamefont {Maeno}, \citenamefont {Beyersdorf}, \citenamefont {Fejer},\ and\ \citenamefont {Kapitulnik}}]{Xia2006-rk}%
  \BibitemOpen
  \bibfield  {author} {\bibinfo {author} {\bibfnamefont {J.}~\bibnamefont {Xia}}, \bibinfo {author} {\bibfnamefont {Y.}~\bibnamefont {Maeno}}, \bibinfo {author} {\bibfnamefont {P.~T.}\ \bibnamefont {Beyersdorf}}, \bibinfo {author} {\bibfnamefont {M.~M.}\ \bibnamefont {Fejer}},\ and\ \bibinfo {author} {\bibfnamefont {A.}~\bibnamefont {Kapitulnik}},\ }\href {https://doi.org/10.1103/PhysRevLett.97.167002} {\bibfield  {journal} {\bibinfo  {journal} {Phys. Rev. Lett.}\ }\textbf {\bibinfo {volume} {97}},\ \bibinfo {pages} {167002} (\bibinfo {year} {2006}{\natexlab{a}})}\BibitemShut {NoStop}%
\bibitem [{\citenamefont {Schemm}\ \emph {et~al.}(2014)\citenamefont {Schemm}, \citenamefont {Gannon}, \citenamefont {Wishne}, \citenamefont {Halperin},\ and\ \citenamefont {Kapitulnik}}]{Schemm2014-cz}%
  \BibitemOpen
  \bibfield  {author} {\bibinfo {author} {\bibfnamefont {E.~R.}\ \bibnamefont {Schemm}}, \bibinfo {author} {\bibfnamefont {W.~J.}\ \bibnamefont {Gannon}}, \bibinfo {author} {\bibfnamefont {C.~M.}\ \bibnamefont {Wishne}}, \bibinfo {author} {\bibfnamefont {W.~P.}\ \bibnamefont {Halperin}},\ and\ \bibinfo {author} {\bibfnamefont {A.}~\bibnamefont {Kapitulnik}},\ }\href {https://doi.org/10.1126/science.1248552} {\bibfield  {journal} {\bibinfo  {journal} {Science}\ }\textbf {\bibinfo {volume} {345}},\ \bibinfo {pages} {190} (\bibinfo {year} {2014})}\BibitemShut {NoStop}%
\bibitem [{\citenamefont {Wei}\ \emph {et~al.}(2022)\citenamefont {Wei}, \citenamefont {Saykin}, \citenamefont {Miller}, \citenamefont {Ran}, \citenamefont {Saha}, \citenamefont {Agterberg}, \citenamefont {Schmalian}, \citenamefont {Butch}, \citenamefont {Paglione},\ and\ \citenamefont {Kapitulnik}}]{Wei2022-hn}%
  \BibitemOpen
  \bibfield  {author} {\bibinfo {author} {\bibfnamefont {D.~S.}\ \bibnamefont {Wei}}, \bibinfo {author} {\bibfnamefont {D.}~\bibnamefont {Saykin}}, \bibinfo {author} {\bibfnamefont {O.~Y.}\ \bibnamefont {Miller}}, \bibinfo {author} {\bibfnamefont {S.}~\bibnamefont {Ran}}, \bibinfo {author} {\bibfnamefont {S.~R.}\ \bibnamefont {Saha}}, \bibinfo {author} {\bibfnamefont {D.~F.}\ \bibnamefont {Agterberg}}, \bibinfo {author} {\bibfnamefont {J.}~\bibnamefont {Schmalian}}, \bibinfo {author} {\bibfnamefont {N.~P.}\ \bibnamefont {Butch}}, \bibinfo {author} {\bibfnamefont {J.}~\bibnamefont {Paglione}},\ and\ \bibinfo {author} {\bibfnamefont {A.}~\bibnamefont {Kapitulnik}},\ }\href {https://doi.org/10.1103/physrevb.105.024521} {\bibfield  {journal} {\bibinfo  {journal} {Phys. Rev. B.}\ }\textbf {\bibinfo {volume} {105}},\ \bibinfo {pages} {024521} (\bibinfo {year} {2022})}\BibitemShut {NoStop}%
\bibitem [{\citenamefont {Ajeesh}\ \emph {et~al.}(2023)\citenamefont {Ajeesh}, \citenamefont {Bordelon}, \citenamefont {Girod}, \citenamefont {Mishra}, \citenamefont {Ronning}, \citenamefont {Bauer}, \citenamefont {Maiorov}, \citenamefont {Thompson}, \citenamefont {Rosa},\ and\ \citenamefont {Thomas}}]{Ajeesh2023-kz}%
  \BibitemOpen
  \bibfield  {author} {\bibinfo {author} {\bibfnamefont {M.~O.}\ \bibnamefont {Ajeesh}}, \bibinfo {author} {\bibfnamefont {M.}~\bibnamefont {Bordelon}}, \bibinfo {author} {\bibfnamefont {C.}~\bibnamefont {Girod}}, \bibinfo {author} {\bibfnamefont {S.}~\bibnamefont {Mishra}}, \bibinfo {author} {\bibfnamefont {F.}~\bibnamefont {Ronning}}, \bibinfo {author} {\bibfnamefont {E.~D.}\ \bibnamefont {Bauer}}, \bibinfo {author} {\bibfnamefont {B.}~\bibnamefont {Maiorov}}, \bibinfo {author} {\bibfnamefont {J.~D.}\ \bibnamefont {Thompson}}, \bibinfo {author} {\bibfnamefont {P.~F.~S.}\ \bibnamefont {Rosa}},\ and\ \bibinfo {author} {\bibfnamefont {S.~M.}\ \bibnamefont {Thomas}},\ }\href {https://doi.org/10.1103/physrevx.13.041019} {\bibfield  {journal} {\bibinfo  {journal} {Phys. Rev. X.}\ }\textbf {\bibinfo {volume} {13}},\ \bibinfo {pages} {041019} (\bibinfo {year} {2023})}\BibitemShut {NoStop}%
\bibitem [{\citenamefont {Ikeda}\ \emph {et~al.}(2026)\citenamefont {Ikeda}, \citenamefont {Nakamura}, \citenamefont {Yamane}, \citenamefont {Noda}, \citenamefont {Ikeda},\ and\ \citenamefont {Yonezawa}}]{Ikeda2026-bq}%
  \BibitemOpen
  \bibfield  {author} {\bibinfo {author} {\bibfnamefont {A.}~\bibnamefont {Ikeda}}, \bibinfo {author} {\bibfnamefont {S.}~\bibnamefont {Nakamura}}, \bibinfo {author} {\bibfnamefont {S.}~\bibnamefont {Yamane}}, \bibinfo {author} {\bibfnamefont {K.}~\bibnamefont {Noda}}, \bibinfo {author} {\bibfnamefont {A.}~\bibnamefont {Ikeda}},\ and\ \bibinfo {author} {\bibfnamefont {S.}~\bibnamefont {Yonezawa}},\ }\href {https://doi.org/10.1103/vy7j-ylb4} {\bibfield  {journal} {\bibinfo  {journal} {Phys. Rev. Res.}\ }\textbf {\bibinfo {volume} {8}},\ \bibinfo {pages} {013169} (\bibinfo {year} {2026})}\BibitemShut {NoStop}%
\bibitem [{\citenamefont {Yamane}\ \emph {et~al.}(2026)\citenamefont {Yamane}, \citenamefont {Nakamura}, \citenamefont {Ikeda}, \citenamefont {Noda}, \citenamefont {Ikeda},\ and\ \citenamefont {Yonezawa}}]{Yamane2026-jl}%
  \BibitemOpen
  \bibfield  {author} {\bibinfo {author} {\bibfnamefont {S.}~\bibnamefont {Yamane}}, \bibinfo {author} {\bibfnamefont {S.}~\bibnamefont {Nakamura}}, \bibinfo {author} {\bibfnamefont {A.}~\bibnamefont {Ikeda}}, \bibinfo {author} {\bibfnamefont {K.}~\bibnamefont {Noda}}, \bibinfo {author} {\bibfnamefont {A.}~\bibnamefont {Ikeda}},\ and\ \bibinfo {author} {\bibfnamefont {S.}~\bibnamefont {Yonezawa}},\ }\href {https://doi.org/10.56646/jjapcp.12.0\_011011} {\bibfield  {journal} {\bibinfo  {journal} {JJAP Conference Proceedings}\ }\textbf {\bibinfo {volume} {12}},\ \bibinfo {pages} {011011} (\bibinfo {year} {2026})}\BibitemShut {NoStop}%
\bibitem [{\citenamefont {Sullivan}\ and\ \citenamefont {Kaszynski}(2019)}]{Sullivan2019-up}%
  \BibitemOpen
  \bibfield  {author} {\bibinfo {author} {\bibfnamefont {C.}~\bibnamefont {Sullivan}}\ and\ \bibinfo {author} {\bibfnamefont {A.}~\bibnamefont {Kaszynski}},\ }\href {https://doi.org/10.21105/joss.01450} {\bibfield  {journal} {\bibinfo  {journal} {J. Open Source Softw.}\ }\textbf {\bibinfo {volume} {4}},\ \bibinfo {pages} {1450} (\bibinfo {year} {2019})}\BibitemShut {NoStop}%
\bibitem [{\citenamefont {Ong}\ \emph {et~al.}(2013)\citenamefont {Ong}, \citenamefont {Richards}, \citenamefont {Jain}, \citenamefont {Hautier}, \citenamefont {Kocher}, \citenamefont {Cholia}, \citenamefont {Gunter}, \citenamefont {Chevrier}, \citenamefont {Persson},\ and\ \citenamefont {Ceder}}]{Ong2013-hn}%
  \BibitemOpen
  \bibfield  {author} {\bibinfo {author} {\bibfnamefont {S.~P.}\ \bibnamefont {Ong}}, \bibinfo {author} {\bibfnamefont {W.~D.}\ \bibnamefont {Richards}}, \bibinfo {author} {\bibfnamefont {A.}~\bibnamefont {Jain}}, \bibinfo {author} {\bibfnamefont {G.}~\bibnamefont {Hautier}}, \bibinfo {author} {\bibfnamefont {M.}~\bibnamefont {Kocher}}, \bibinfo {author} {\bibfnamefont {S.}~\bibnamefont {Cholia}}, \bibinfo {author} {\bibfnamefont {D.}~\bibnamefont {Gunter}}, \bibinfo {author} {\bibfnamefont {V.~L.}\ \bibnamefont {Chevrier}}, \bibinfo {author} {\bibfnamefont {K.~A.}\ \bibnamefont {Persson}},\ and\ \bibinfo {author} {\bibfnamefont {G.}~\bibnamefont {Ceder}},\ }\href {https://doi.org/10.1016/j.commatsci.2012.10.028} {\bibfield  {journal} {\bibinfo  {journal} {Comput. Mater. Sci.}\ }\textbf {\bibinfo {volume} {68}},\ \bibinfo {pages} {314} (\bibinfo {year} {2013})}\BibitemShut {NoStop}%
\bibitem [{\citenamefont {Norton}(1989)}]{Norton1989-ye}%
  \BibitemOpen
  \bibfield  {author} {\bibinfo {author} {\bibfnamefont {M.~L.}\ \bibnamefont {Norton}},\ }\href {https://doi.org/10.1016/0025-5408(89)90145-1} {\bibfield  {journal} {\bibinfo  {journal} {Mater. Res. Bull.}\ }\textbf {\bibinfo {volume} {24}},\ \bibinfo {pages} {1391} (\bibinfo {year} {1989})}\BibitemShut {NoStop}%
\bibitem [{\citenamefont {Nishio}\ \emph {et~al.}(2001)\citenamefont {Nishio}, \citenamefont {Minami},\ and\ \citenamefont {Uwe}}]{Nishio2001-xg}%
  \BibitemOpen
  \bibfield  {author} {\bibinfo {author} {\bibfnamefont {T.}~\bibnamefont {Nishio}}, \bibinfo {author} {\bibfnamefont {H.}~\bibnamefont {Minami}},\ and\ \bibinfo {author} {\bibfnamefont {H.}~\bibnamefont {Uwe}},\ }\href {https://doi.org/10.1016/s0921-4534(01)00255-6} {\bibfield  {journal} {\bibinfo  {journal} {Physica C Supercond.}\ }\textbf {\bibinfo {volume} {357-360}},\ \bibinfo {pages} {376} (\bibinfo {year} {2001})}\BibitemShut {NoStop}%
\bibitem [{\citenamefont {Aharoni}(1998)}]{Aharoni1998-sx}%
  \BibitemOpen
  \bibfield  {author} {\bibinfo {author} {\bibfnamefont {A.}~\bibnamefont {Aharoni}},\ }\href {https://doi.org/10.1063/1.367113} {\bibfield  {journal} {\bibinfo  {journal} {J. Appl. Phys.}\ }\textbf {\bibinfo {volume} {83}},\ \bibinfo {pages} {3432} (\bibinfo {year} {1998})}\BibitemShut {NoStop}%
\bibitem [{\citenamefont {Pei}\ \emph {et~al.}(1990)\citenamefont {Pei}, \citenamefont {Jorgensen}, \citenamefont {Dabrowski}, \citenamefont {Hinks}, \citenamefont {Richards}, \citenamefont {Mitchell}, \citenamefont {Newsam}, \citenamefont {Sinha}, \citenamefont {Vaknin},\ and\ \citenamefont {Jacobson}}]{Pei1990-xo}%
  \BibitemOpen
  \bibfield  {author} {\bibinfo {author} {\bibfnamefont {S.}~\bibnamefont {Pei}}, \bibinfo {author} {\bibfnamefont {J.~D.}\ \bibnamefont {Jorgensen}}, \bibinfo {author} {\bibfnamefont {B.}~\bibnamefont {Dabrowski}}, \bibinfo {author} {\bibfnamefont {D.~G.}\ \bibnamefont {Hinks}}, \bibinfo {author} {\bibfnamefont {D.~R.}\ \bibnamefont {Richards}}, \bibinfo {author} {\bibfnamefont {A.~W.}\ \bibnamefont {Mitchell}}, \bibinfo {author} {\bibfnamefont {J.~M.}\ \bibnamefont {Newsam}}, \bibinfo {author} {\bibfnamefont {S.~K.}\ \bibnamefont {Sinha}}, \bibinfo {author} {\bibfnamefont {D.}~\bibnamefont {Vaknin}},\ and\ \bibinfo {author} {\bibfnamefont {A.~J.}\ \bibnamefont {Jacobson}},\ }\href {https://doi.org/10.1103/physrevb.41.4126} {\bibfield  {journal} {\bibinfo  {journal} {Phys. Rev. B Condens. Matter}\ }\textbf {\bibinfo {volume} {41}},\ \bibinfo {pages} {4126} (\bibinfo {year} {1990})}\BibitemShut {NoStop}%
\bibitem [{\citenamefont {Kim}\ \emph {et~al.}(2000)\citenamefont {Kim}, \citenamefont {Kang}, \citenamefont {Kim}, \citenamefont {Kim}, \citenamefont {Schmidbauer},\ and\ \citenamefont {Hodby}}]{Kim2000-sy}%
  \BibitemOpen
  \bibfield  {author} {\bibinfo {author} {\bibfnamefont {H.-T.}\ \bibnamefont {Kim}}, \bibinfo {author} {\bibfnamefont {K.-Y.}\ \bibnamefont {Kang}}, \bibinfo {author} {\bibfnamefont {B.-J.}\ \bibnamefont {Kim}}, \bibinfo {author} {\bibfnamefont {Y.~C.}\ \bibnamefont {Kim}}, \bibinfo {author} {\bibfnamefont {W.}~\bibnamefont {Schmidbauer}},\ and\ \bibinfo {author} {\bibfnamefont {J.~W.}\ \bibnamefont {Hodby}},\ }\href {https://doi.org/10.1016/s0921-4534(00)00664-x} {\bibfield  {journal} {\bibinfo  {journal} {Physica C Supercond.}\ }\textbf {\bibinfo {volume} {341-348}},\ \bibinfo {pages} {729} (\bibinfo {year} {2000})}\BibitemShut {NoStop}%
\bibitem [{\citenamefont {Tao}\ \emph {et~al.}(2015)\citenamefont {Tao}, \citenamefont {Deng}, \citenamefont {Yang}, \citenamefont {Wang}, \citenamefont {Zhu},\ and\ \citenamefont {Wen}}]{Tao2015-sm}%
  \BibitemOpen
  \bibfield  {author} {\bibinfo {author} {\bibfnamefont {J.}~\bibnamefont {Tao}}, \bibinfo {author} {\bibfnamefont {Q.}~\bibnamefont {Deng}}, \bibinfo {author} {\bibfnamefont {H.}~\bibnamefont {Yang}}, \bibinfo {author} {\bibfnamefont {Z.}~\bibnamefont {Wang}}, \bibinfo {author} {\bibfnamefont {X.}~\bibnamefont {Zhu}},\ and\ \bibinfo {author} {\bibfnamefont {H.-H.}\ \bibnamefont {Wen}},\ }\href {https://doi.org/10.1103/PhysRevB.91.214516} {\bibfield  {journal} {\bibinfo  {journal} {Phys. Rev. B}\ }\textbf {\bibinfo {volume} {91}},\ \bibinfo {pages} {214516} (\bibinfo {year} {2015})}\BibitemShut {NoStop}%
\bibitem [{\citenamefont {Xia}\ \emph {et~al.}(2006{\natexlab{b}})\citenamefont {Xia}, \citenamefont {Beyersdorf}, \citenamefont {Fejer},\ and\ \citenamefont {Kapitulnik}}]{Xia2006-sg}%
  \BibitemOpen
  \bibfield  {author} {\bibinfo {author} {\bibfnamefont {J.}~\bibnamefont {Xia}}, \bibinfo {author} {\bibfnamefont {P.~T.}\ \bibnamefont {Beyersdorf}}, \bibinfo {author} {\bibfnamefont {M.~M.}\ \bibnamefont {Fejer}},\ and\ \bibinfo {author} {\bibfnamefont {A.}~\bibnamefont {Kapitulnik}},\ }\href {https://doi.org/10.1063/1.2336620} {\bibfield  {journal} {\bibinfo  {journal} {Appl. Phys. Lett.}\ }\textbf {\bibinfo {volume} {89}},\ \bibinfo {pages} {062508} (\bibinfo {year} {2006}{\natexlab{b}})}\BibitemShut {NoStop}%
\bibitem [{\citenamefont {Buchner}\ \emph {et~al.}(2018)\citenamefont {Buchner}, \citenamefont {H{\"{o}}fler}, \citenamefont {Henne}, \citenamefont {Ney},\ and\ \citenamefont {Ney}}]{Buchner2018-xm}%
  \BibitemOpen
  \bibfield  {author} {\bibinfo {author} {\bibfnamefont {M.}~\bibnamefont {Buchner}}, \bibinfo {author} {\bibfnamefont {K.}~\bibnamefont {H{\"{o}}fler}}, \bibinfo {author} {\bibfnamefont {B.}~\bibnamefont {Henne}}, \bibinfo {author} {\bibfnamefont {V.}~\bibnamefont {Ney}},\ and\ \bibinfo {author} {\bibfnamefont {A.}~\bibnamefont {Ney}},\ }\href {https://doi.org/10.1063/1.5045299} {\bibfield  {journal} {\bibinfo  {journal} {J. Appl. Phys.}\ }\textbf {\bibinfo {volume} {124}},\ \bibinfo {pages} {161101} (\bibinfo {year} {2018})}\BibitemShut {NoStop}%
\bibitem [{\citenamefont {\v{S}im\v{s}a}\ \emph {et~al.}(1980)\citenamefont {\v{S}im\v{s}a}, \citenamefont {Legall},\ and\ \citenamefont {{\v{S}}irok\'{y}}}]{Simsa1980-us}%
  \BibitemOpen
  \bibfield  {author} {\bibinfo {author} {\bibfnamefont {Z.}~\bibnamefont {\v{S}im\v{s}a}}, \bibinfo {author} {\bibfnamefont {H.}~\bibnamefont {Legall}},\ and\ \bibinfo {author} {\bibfnamefont {P.}~\bibnamefont {{\v{S}}irok\'{y}}},\ }\href {https://doi.org/10.1002/pssb.2221000234} {\bibfield  {journal} {\bibinfo  {journal} {Phys. Stat. Solidi B Basic Res.}\ }\textbf {\bibinfo {volume} {100}},\ \bibinfo {pages} {665} (\bibinfo {year} {1980})}\BibitemShut {NoStop}%
\bibitem [{\citenamefont {Fontijn}\ \emph {et~al.}(1997)\citenamefont {Fontijn}, \citenamefont {van~der Zaag}, \citenamefont {Devillers}, \citenamefont {Brabers},\ and\ \citenamefont {Metselaar}}]{Fontijn1997-hr}%
  \BibitemOpen
  \bibfield  {author} {\bibinfo {author} {\bibfnamefont {W.~F.~J.}\ \bibnamefont {Fontijn}}, \bibinfo {author} {\bibfnamefont {P.~J.}\ \bibnamefont {van~der Zaag}}, \bibinfo {author} {\bibfnamefont {M.~A.~C.}\ \bibnamefont {Devillers}}, \bibinfo {author} {\bibfnamefont {V.~A.~M.}\ \bibnamefont {Brabers}},\ and\ \bibinfo {author} {\bibfnamefont {R.}~\bibnamefont {Metselaar}},\ }\href {https://doi.org/10.1103/physrevb.56.5432} {\bibfield  {journal} {\bibinfo  {journal} {Phys. Rev. B Condens. Matter}\ }\textbf {\bibinfo {volume} {56}},\ \bibinfo {pages} {5432} (\bibinfo {year} {1997})}\BibitemShut {NoStop}%
\bibitem [{\citenamefont {Kim}\ \emph {et~al.}(1963)\citenamefont {Kim}, \citenamefont {Hempstead},\ and\ \citenamefont {Strnad}}]{Kim1963-xe}%
  \BibitemOpen
  \bibfield  {author} {\bibinfo {author} {\bibfnamefont {Y.~B.}\ \bibnamefont {Kim}}, \bibinfo {author} {\bibfnamefont {C.~F.}\ \bibnamefont {Hempstead}},\ and\ \bibinfo {author} {\bibfnamefont {A.~R.}\ \bibnamefont {Strnad}},\ }\href {https://doi.org/10.1103/physrev.129.528} {\bibfield  {journal} {\bibinfo  {journal} {Phys. Rev.}\ }\textbf {\bibinfo {volume} {129}},\ \bibinfo {pages} {528} (\bibinfo {year} {1963})}\BibitemShut {NoStop}%
\bibitem [{\citenamefont {Chen}\ \emph {et~al.}(1990)\citenamefont {Chen}, \citenamefont {Sanchez}, \citenamefont {Nogues},\ and\ \citenamefont {Muoz}}]{Chen1990-yo}%
  \BibitemOpen
  \bibfield  {author} {\bibinfo {author} {\bibfnamefont {D.}~\bibnamefont {Chen}}, \bibinfo {author} {\bibfnamefont {A.}~\bibnamefont {Sanchez}}, \bibinfo {author} {\bibfnamefont {J.}~\bibnamefont {Nogues}},\ and\ \bibinfo {author} {\bibfnamefont {J.~S.}\ \bibnamefont {Muoz}},\ }\href {https://doi.org/10.1103/physrevb.41.9510} {\bibfield  {journal} {\bibinfo  {journal} {Phys. Rev. B Condens. Matter}\ }\textbf {\bibinfo {volume} {41}},\ \bibinfo {pages} {9510} (\bibinfo {year} {1990})}\BibitemShut {NoStop}%
\bibitem [{\citenamefont {Barilo}\ \emph {et~al.}(1995)\citenamefont {Barilo}, \citenamefont {Gatalskaya}, \citenamefont {Shiryaev}, \citenamefont {Shestac}, \citenamefont {Kurochkin}, \citenamefont {Smirnova}, \citenamefont {Koyava}, \citenamefont {Orlova},\ and\ \citenamefont {Pushkarev}}]{Barilo1995-my}%
  \BibitemOpen
  \bibfield  {author} {\bibinfo {author} {\bibfnamefont {S.~N.}\ \bibnamefont {Barilo}}, \bibinfo {author} {\bibfnamefont {V.~I.}\ \bibnamefont {Gatalskaya}}, \bibinfo {author} {\bibfnamefont {S.~V.}\ \bibnamefont {Shiryaev}}, \bibinfo {author} {\bibfnamefont {A.~S.}\ \bibnamefont {Shestac}}, \bibinfo {author} {\bibfnamefont {L.~A.}\ \bibnamefont {Kurochkin}}, \bibinfo {author} {\bibfnamefont {T.~V.}\ \bibnamefont {Smirnova}}, \bibinfo {author} {\bibfnamefont {V.~T.}\ \bibnamefont {Koyava}}, \bibinfo {author} {\bibfnamefont {N.~S.}\ \bibnamefont {Orlova}},\ and\ \bibinfo {author} {\bibfnamefont {A.~V.}\ \bibnamefont {Pushkarev}},\ }\href {https://doi.org/10.1016/0921-4534(95)00556-0} {\bibfield  {journal} {\bibinfo  {journal} {Physica C Supercond.}\ }\textbf {\bibinfo {volume} {254}},\ \bibinfo {pages} {181} (\bibinfo {year} {1995})}\BibitemShut {NoStop}%
\bibitem [{\citenamefont {Schemm}\ \emph {et~al.}(2015)\citenamefont {Schemm}, \citenamefont {Baumbach}, \citenamefont {Tobash}, \citenamefont {Ronning}, \citenamefont {Bauer},\ and\ \citenamefont {Kapitulnik}}]{Schemm2015-tr}%
  \BibitemOpen
  \bibfield  {author} {\bibinfo {author} {\bibfnamefont {E.~R.}\ \bibnamefont {Schemm}}, \bibinfo {author} {\bibfnamefont {R.~E.}\ \bibnamefont {Baumbach}}, \bibinfo {author} {\bibfnamefont {P.~H.}\ \bibnamefont {Tobash}}, \bibinfo {author} {\bibfnamefont {F.}~\bibnamefont {Ronning}}, \bibinfo {author} {\bibfnamefont {E.~D.}\ \bibnamefont {Bauer}},\ and\ \bibinfo {author} {\bibfnamefont {A.}~\bibnamefont {Kapitulnik}},\ }\href {https://doi.org/10.1103/physrevb.91.140506} {\bibfield  {journal} {\bibinfo  {journal} {Phys. Rev. B Condens. Matter Mater. Phys.}\ }\textbf {\bibinfo {volume} {91}},\ \bibinfo {pages} {140506} (\bibinfo {year} {2015})}\BibitemShut {NoStop}%
\end{thebibliography}%

\end{document}